\documentclass[a4paper,aip,jmp,reprint,onecolumn,author-year]{revtex4-1}

\usepackage[latin1]{inputenc}      
\usepackage[T1]{fontenc}
\usepackage{ae,aecompl}     
\usepackage[english]{babel}

\usepackage[newenum,newitem]{paralist}
\usepackage[tight,nice]{units}

\usepackage{graphicx}  
\usepackage{dcolumn}   
\usepackage{enumerate}

\usepackage{pictexwd,dcpic}  

\usepackage{rmath}
\usepackage{rtheorems} 
\usepackage{color}
\usepackage{ifthen}
\definecolor{Rot}{rgb}{1.0,0,0}
\def\todo#1{}

\def\version{\texttt{v2.3}}

\usepackage{relsize}
\usepackage{scrpage2}
\deftripstyle{default}{\shorttitle}{}{\thepage}{}{}{%
\normalfont\smaller\version}

\def\h#1{\f{#1}{2}}
\def\minsec#1{\bigskip\noindent\textbf{#1}\\[-2mm]}

\def\BibitemShut#1{}

\DeclareMathAlphabet{\mathitbf}{OML}{cmm}{b}{it}

\def\symtens{\vee}
\def\Clalg{\Cl^c_{1,3}}
\def\C{\set C}
\def\R{\set R}
\def\repr#1#2{D^{(#1,#2)}}
\def\pauli{\tilde\sigma}
\def\dirmat{\tilde\gamma}

\def\complexi{\mathitbf{i}} 

\def\nocontentsline#1#2#3{}
\let\addcontentslineold=\addcontentsline 
\def\hiddensubsection#1{%
   \let\addcontentsline=\nocontentsline%
   \subsection{#1}%
   \let\addcontentsline=\addcontentslineold%
}
\def\acknowledgements{%
   \let\addcontentsline=\nocontentsline%
   \section*{Acknowledgements}%
   \let\addcontentsline=\addcontentslineold%
}

\begin{document}


\def\shorttitle{Rainer M\"{u}hlhoff -- Higher Spin Quantum Fields as Twisted
Dirac Fields}
\title[\shorttitle]{Higher Spin Quantum Fields as Twisted Dirac Fields}


\author{Rainer M\"{u}hlhoff}%
\email{rainer.muehlhoff@itp.uni-leipzig.de}
\affiliation{Institut für Theoretische Physik, Universität Leipzig, 
Postfach 100\,920, D-04009 Leipzig, Germany}

\begin{abstract}
	\textbf{Abstract.} 
	We pursue the idea of constructing higher spin fields as solutions to
	twisted Dirac operators. As general results we find that twisted
	prenormally hyperbolic first order operators (such as the Dirac
	operator) on globally hyperbolic Lorentzian spacetime manifolds have
	unique advanced and retarded Green's functions and their Cauchy
	problem is well-posed. The space of compactly supported Cauchy data
	can be equipped with a naturally induced Hermitian but generally
	indefinite product, which is shown to be independent of the underlying
	choice of Cauchy hypersurface. 
	
	As a special result, we study one particular example of twisted Dirac
	operator for Fermionic fields of higher spin. It allows a canonical
	Hermitian product on the spinor bundle, but the resulting product on
	the space of compactly supported Cauchy data fails positivity. 
	Thus, it is shown that this construction does not allow quantization 
	by forming a $C^*$-algebra representation of the canonical
	anti-commutation relations in the well known fashion generalizing the
	spin $\halb$ construction by \cite{Dimock1982}. 

	This document is equipped with an introduction to the
	formalism of 2-spinors on curved spacetimes in an invariant fashion
	using abstract index notation. 
\end{abstract}

\maketitle

\vspace{-5mm}
\tableofcontents

\vspace{12mm}

\section{Introduction and Motivation}
\label{sec:history}

\noindent 
In the setting of general relativity, various types of physical fields (and,
in a suitable sense, also quantized fields) can be described as solutions to a
differential equation 
\[
\mathcal{T}\Phi + \complexi\, m\,\Phi = 0
\,,\qquad \Phi\in\Gamma(\mathscr{E}) \,,\ m\in\set R
\]. 
Here, $\mathscr{E}$ is a real or complex vector bundle over the Lorentzian
spacetime manifold $M$ with metric $g$, and $\mathcal{T}$ is a linear
differential operator while $\Gamma(\mathscr{E})$ denotes the space of smooth
sections of $\mathscr{E}$. $m$ is the physical mass and may or may not be
vanishing. 

Based on previous investigations on Buchdahl's higher spin equations (cf.\
\onlinecite{Dipl}), this work started from the idea of constructing
\textbf{higher spin field equations as twisted Dirac operators.} The Dirac
operator is the
most prominent representative of the class of prenormally hyperbolic first
order differential operators, which were shown to have unique advanced and
retarded Green's functions and a well-posed Cauchy problem on globally
hyperbolic Lorentzian spacetime manifolds in \cite{Mueh2011a}. 

In this paper we shall show: The twist of a prenormally hyperbolic operator
remains prenormally hyperbolic. And if there is a Hermitian metric on the
twisted bundle, it also allows a natural \textbf{construction of Hermitian
product} on the vector space of compactly supported Cauchy initial data, which
is \textbf{independent of the choice of Cauchy hypersurface} on which the
initial data are located. So far we are generalizing the well-known spin
$\halb$ construction by \cite{Dimock1982}. However, unlike in the case of spin
$\halb$, the Hermitian product on the space of compactly supported Cauchy data
will in general not be positive-definite, so that in general there is no
resulting pre-Hilbert space structure and a $\CAR$-algebraic quantization
construction will not be immediate (cf.\ \onlinecite[sec.\ V]{Mueh2011a} for a
brief outline of the quantization construction using $\CAR$-algebras. Cf.\
also the recent preprint \onlinecite{BaerGin2011} for related results and
algebraic quantization in a more general setting.)

Moreover, in this paper we will single out one particular twisted Dirac
operator for
\textbf{Fermionic fields,} which allows a canonical Hermitian bundle product
which comes from a \textbf{generalized Dirac adjoint} on the higher spin
bundle. However, these operators will be of indefinite type with respect to
this bundle product, so that a $\CAR$-algebraic quantization construction in
the style of \cite{Dimock1982} turns out to fail. 


\bigskip\noindent
The search for a differential operator on higher spin fields has a long
history. 
\cite{Dirac1936} himself presented such attempts in flat spacetime but which
were found inconsistent (i.\,e.\ integrable only under unacceptably strong
assumptions) in the setting of minimal coupling to an
electromagnetic field if spin $>1$ by \cite{FierzPauli1939}. 
Minimal coupling to gravity (by naively replacing partial derivatives by
covariant derivatives on a curved spacetime background) was studied by
\cite{Buchdahl1962}. By adding to the operator a correction term of order 0 in
\nocite{Buchdahl1982a}\nocite{Buchdahl1982b}%
Buchdahl (1982a, 1982b)%
, he was able to make the 
minimally coupled \textit{massive} equations integrable for arbitrary spin
under certain \textit{constraint conditions} 
(for a review and additional considerations in modern language, cf.\
\onlinecite{Dipl}). 
\cite{Wuensch1985} and 
\nocite{Illge1993}\nocite{Illge1996}%
Illge (1993, 1996), 
\cite{IllgeSchimming1999} also worked on Buchdahl's equations, but -- though
they are consistent (with constraints) -- a natural construction of
Hermitian scalar product on the respective spinor bundle
which yields a pre-Hilbert space structure on the space of solutions to the
Cauchy problem with compactly supported initial data, could not be established. 

\bigskip\noindent 
Dealing with spinor fields on manifolds requires a sophisticated piece of
machinery in terms of notation and geometric formalism. To make this document
accessible to both theoretical physicists accustomed to index notation as well
as to mathematicians with a background in spin geometry, as a general
principle every central result shall be stated both in an index- as well as in
a mathematical (non-index) notation. In proofs and calculations though 
we shall often completely rely on index formalism where this brings significant
notational benefits. 

But anyway, the index formalism adopted in this document is an \textbf{abstract
index notation.} All expressions (with or without indices) are invariant
if not otherwise stated (cf.\ also the remarks in appendix
\ref{sec:app:index-notation}). To make this document more readable to a wider
audience, there is a compact but solid introduction to spinor formalism in the
appendix.

\section{Notation and conventions}
\label{sec:notation}

\subsection{Conventions} 

\noindent 
\textbf{Geometric background:} 
Throughout the document, let $(M,g)$ be a time-oriented and space-oriented,
4-dimensional, globally hyperbolic, Lorentzian manifold of signature
\mbox{$(+---)$}.
In particular, this implies that $(M,g)$ is oriented and connected, satisfies
the strong causality condition and has a spin structure (more details are
summarized in \cite{Dipl}; for the concept of global hyperbolicity cf.\
\cite{BaerGinPfae}, existence of a spin structure was shown in 
\onlinecite{Geroch70b}). 
We shall refer to $(M,g)$ as \textbf{spacetime manifold.} 

\textbf{Vector bundles:} 
For a vector bundle $\mathscr{E}$ on $M$, $\Gamma(\mathscr{E})$ is the space
of $C^\infty$- (i.\,e.\ smooth) sections, $\Gamma_0(\mathscr{E})$ the space of
compactly supported $C^\infty$-sections. $\mathscr{E}^*$ denotes the dual
bundle, $\End({\mathscr{E}})$ the bundle of endomorphisms and
$\GL(\mathscr{E})$ the bundle of automorphisms of $\mathscr{E}$. 
By $\Id_{\mathscr{E}}\in \Gamma(\GL(\mathscr{E}))$ we denote the identity map
on the fibers. 
For a point $m\in M$, the fiber of $\mathscr{E}$ over $m$ is denoted by
$\mathscr{E}_m$. 
A covariant derivatives on $\mathscr{E}$ will be denoted by
$\nabla^\mathscr{E}$ if it is important to distinguish it from other covariant
derivatives. The \textbf{tangent bundle} on $(M,g)$ will be
denoted $TM$, the co-tangent bundle $T^*M$. We always assume $TM$ and $T^*M$ to
be equipped with the metric induced \textbf{Levi-Civita covariant derivative,}
denoted by $\nabla$. 

\textbf{Indices:} Indices for vectors, tensors and spinors are denoted as
superscripts, indices for co-vectors, co-tensors and co-spinors as subscripts.
Latin lower case letters $a, b, c, \ldots$ from the beginning of the alphabet
denote abstract \textbf{(co-)vector/tensor indices.} 
The subscript indices $i,k,l$ are used as numerical counting indices
running from 1 (i.\,e., they take values $1, 2, 3, \ldots$), e.\,g.\ the
$i$-th Pauli spin matrix will be denoted $\pauli_i$, for $i=1,2,3$. To
distinguish the counting index $i$ from the imaginary number $\complexi\in\C$, the
latter is printed in bold face.  Latin capital letters $A,B,C,\ldots,$ denote
abstract indices of \textbf{positive Weyl (co-)spinors,} dotted Latin capital
letters $\dot X,\dot Y, \ldots$ denote abstract indices of \textbf{negative
Weyl (co-)spinors} (for an introduction to 2-spinor index notation, cf.\
appendix \ref{sec:app:index-notation}). Finally, Greek lower case letters
$\kappa,\lambda,\mu,\nu, \ldots$ are classical \textbf{spacetime component
indices,} usually taking values from 0 to 3, i.\,e., $\mu=0,\ldots,3$.
\textbf{Implicit contraction} (in the case of abstract indices) and
\textbf{implicit summation} (in the case of classical spacetime component
indices) is taken on \textit{diagonal} pairs of similar indices, e.\,g.\ for
$X=X^a \in \Gamma(TM)$, $\varphi=\varphi_a\in \Gamma(T^*M)$, we write
$\varphi(X) = \varphi_aX^a$. \textbf{Index shifting} of vector/tensor indices
is with respect to the metric $g_{ab}$ on $TM$ and the induced \textbf{inverse
metric} $g^{ab}$ on $T^*M$. Index shifting of spinor indices is with
respect to the $\varepsilon$-spinors $\varepsilon_{AB}$, $\varepsilon^{AB}$,
$\varepsilon_{\dot X\dot Y}$, $\varepsilon^{\dot X\dot Y}$, as described in
appendix \ref{sec:app:index-notation}. 

\textbf{Lorentz geometry:} 
For $m\in M$, a vector $x\in T_mM$ will be called \textbf{timelike}, if
$g(x,x) > 0$ and \textbf{causal} if $g(x,x)\ge 0$. Let the time-orientation of
$(M,g)$ (which was assumed to be chosen above) be represented by a global
timelike vector field $\tau\in\Gamma(TM)$. Then we call $x$ \textbf{future
pointing} or future directed, if $g(x,\tau) > 0$, \textbf{past directed} if
$g(x,\tau) < 0$. A smooth curve $c\: (-\varepsilon,\varepsilon)\to M$ in $M$
is called future directed, past directed, timelike, causal, respectively, if
its tangent vector $\dot c(t)$ has the respective property for all
$t\in(-\varepsilon,\varepsilon)$. Finally, 
for $A\subseteq M$, $J_+(A)$ resp.\ $J_-(A)$ denotes the
\textbf{causal future resp.\ past} of the set $A$, i.\,e.\ the set of points
in $M$, which are either points in $A$ or which can be reached from a point in
$A$ by a future resp.\ past directed, causal, piecewise $C^1$ curve. Moreover,
we set $J(A) := J_+(A)\cup J_-(A)$.

\subsection{Spinors on curved spacetime} 
\label{sec:notation:spinors}

\noindent 
In this section we declare the necessary geometric structures,
in particular different types of spinor bundles, on our 
spacetime manifold $(M,g)$. We assume the reader to be familiar with $\SLtwo$
spinor formalism on Minkowski vector space (i.\,e.\ ``on the fiber''), an
introduction to which may be found in the appendix. 

\begin{enumerate}
	\item Let $\mathscr{F}_c(M,g)$ denote the bundle of oriented and
		time-oriented pseudo-orthogonal frames on $(M,g)$
		\textbf{(connected frame bundle);} it is an $SO^+(1,3) =
		\restlor$-principal bundle. Let a \textbf{spin structure}
		$\Lambda\: \mathscr{S}(M,g)\to \mathscr{F}_c(M,g)$ be chosen,
		i.\,e.\ a $ \Spin_0(1,3)\isomorph \SLtwo$-principal bundle
		$\mathscr{S}(M,g)$ together with a bundle homomorphism
		$\Lambda$ which is in each fiber the 2-1 universal covering
		map $\lambda\: \SLtwo\to\restlor$ (cf.\ appendix
		\ref{sec:app:sigma-tensor-spinor}). 

	\item Let $D\: \SLtwo\to\GL(\Delta)$ be a finite dimensional complex
		representation of $\SLtwo$ (i.\,e., a spinor representation).
		Then the \textbf{bundle of $D$-spinors on $M$} is the
		associated vector bundle 
		\[
		\mathscr{D} := \mathscr{S}(M,g) \times_D \Delta
		\], 
		i.\,e.\ the vector bundle whose fiber at $m\in M$ consists of
		the orbits 
		\[! \tag{O}
		[F_m, \psi]_D := \{ (F_m \cdot S^{-1}, D(S)\,\psi)\| S\in\SLtwo \} 
		\], 
		for $F_m\in\mathscr{S}(M,g)_m$ and $\psi\in\Delta$. 
		Here, $\cdot\,S^{-1}$ denotes the right action of $S^{-1}$ on
		the fiber $\mathscr{S}(M,g)_m$. 
		Thus, for the positive resp.\ negative Weyl spinor
		representations $D^{(\halb,0)}$
		resp.\ $D^{(0,\halb)}$ on $\Delta_{\halb,0}$ resp.\
		$\Delta_{0,\halb}$ 
		(cf.\ appendix \ref{sec:app:classification}), we declare the
		\textbf{bundles of positive/negative Weyl spinors on $(M,g)$,}
		\[
		\mathscr{D}^{(\halb,0)} 
		:= \mathscr{S}(M,g) \times_{\repr{\halb}{0}}
		\Delta_{\halb,0}
		\,,\qquad
		\mathscr{D}^{(0,\halb)} 
		:= \mathscr{S}(M,g) \times_{\repr{0}{\halb}}
		\Delta_{0,\halb}
		\]. 
		Then the \textbf{bundle of Dirac spinors on $(M,g)$} is
		declared as
		\[
		\mathscr{D}^D 
		:= \mathscr{D}^{\halb,0} \oplus (\mathscr{D}^{0,\halb})^*
		\]. 
		Moreover, we shall often be using the \textbf{totally symmetric
		bundles} 
		\[
		\mathscr{\tilde D}^{(\h{k},\h{l})} 
		:= \mathscr{S}(M,g)
		\times_{\tilde D^{(\h{k},\h{l}}} \tilde\Delta_{\h{k},\h{l}}
		\,,\qquad 
		k,l\in\set N
		\], 
		for the representations
		\[
		\tilde D^{(\h{k},\h{l})} 
		:= D^{(\h{k},0)} \tens \big(D^{(0,\h{l})} \big)^* 
		= \big(D^{(\halb,0)} \big)^{\vee k} \tens \big( \big(
		D^{(0,\halb)} \big)^* \big)^{\vee l}
		\], 
		where $\,^{\vee k}$ denotes the $k$-fold symmetrized tensor
		product. Notice, $\repr{\h{k}}{\h{l}} \isomorph
		\tilde D^{(\h{k},\h{l})}$, but these representations do not
		equal (the first has the dotted indices as spinor, the
		other as co-spinor indices; cf.\ appendix
		\ref{sec:app:classification}). 

		Notice, it is easily seen that forming direct sums, tensor
		products, complex conjugates and duals of the spinor bundles
		is compatible with first performing the analog operations on
		the level of representations and then building the associated
		vector bundles. For example: 
		\[
		\mathscr{D}^{\h{k},0} \vee \mathscr{D}^{\h{k'},0} 
		= \mathscr{D}^{\h{k+k'},0}
		\,,\qquad 
		(\mathscr{D}^D)^* =
		\mathscr{S}(M,g)\times_{(D^D)^*}
		\big( \Delta_{\halb,0} \oplus \Delta_{0,\halb}^* \big)^*
		\], 
		where $\vee$ denotes the symmetrized tensor product. 

	\item In analogy to the fiber objects $\varepsilon^{AB}$, $\sigma_a^{\
		A\dot X}$ and $\gamma_a$ (cf.\ appendices
		\ref{sec:app:index-notation},
		\ref{sec:app:sigma-tensor-spinor},
		\ref{sec:app:gamma-tensor-spinor}), we shall now declare 
		objects 
		\[
		\hat\varepsilon^{AB}\in \Gamma(\mathscr{D}^{\halb,0}\tens
		\mathscr{D}^{\halb,0}) 
		\,,\quad 
		\hat\sigma_a^{\ A\dot X} \in \Gamma( T^*M \tens
		(\mathscr{D}^{\halb,\halb})^*) 
		\,,\quad 
		\hat\gamma_a \in \Gamma(T^*M\tens \GL(\mathscr{D}^D))
		\]
		on $(M,g)$. 
		To this end, let first $F$ be a local section of
		$\mathscr{S}(M,g)$ on an open neighborhood $U\subset M$. By
		$\Lambda$, $F$ gets mapped to a local section of 
		$\mathscr{F}_c(M,g)$, which constitutes a local pseudo-orthogonal
		tetrad field $(\hat b_0,\hat b_1, \hat b_2, \hat b_3) :=
		\Lambda\circ F$. 
		Moreover, let $(E_1, E_2)$ denote the standard basis of
		$\Delta_{\halb,0} = \C^2$. Then $F$ induces
		local sections $\hat E_1, \hat E_2$ of $\mathscr{D}^{\halb,0}$ 
		by setting 
		\[
		\hat E_i := [F, E_i]_{D^{(\halb,0)}}
		\,,\quad i=1,2
		\]. 
		
		Now, let the totally anti-symmetric spinor
		$\varepsilon^{AB}\in\Delta_{\halb,0}\tens \Delta_{\halb,0}$ from
		appendix \ref{sec:app:index-notation} be given in components
		$\varepsilon^{\kappa\lambda}$ with respect to the 
		$(E_1,E_2)$. Then we define the spinor field
		$\hat\varepsilon^{AB}$ locally by
		\[
		\hat \varepsilon^{AB} = \varepsilon^{\kappa\lambda}\, \hat
		E_\kappa\tens \hat E_\lambda 
		\,,\qquad \kappa,\lambda=1,2
		\]. 
		(In the same way, we declare $\varepsilon_{AB}$; then we
		obtain $\varepsilon^{\dot X\dot Y}$ and $\varepsilon_{\dot
		X\dot Y}$ by complex conjugation.) 
		Analogously, let the $\sigma$-spinor-tensor $\sigma_a^{\ A\dot
		X}$ from appendix \ref{sec:app:sigma-tensor-spinor} be given
		in component representation $\sigma_\mu^{\
		\kappa\dot\lambda}$ with respect to the standard bases
		$(e^0,\ldots,e^3)$ of $(\R^4,\eta)^*_\C$, $(E_1,E_2)$ of
		$\Delta_{\halb,0}$ and $(\bar E_1,\bar E_2)$ of
		$\Delta_{0,\halb} = \overline{\Delta_{\halb,0}}$. Then we
		define the $\sigma$-tensor spinor field $\hat\sigma_a^{\ A\dot
		X}$ locally by 
		\[
		\hat \sigma_a^{\ A\dot X} 
		= \sigma_\mu^{\ \kappa\dot\lambda}\, \hat
		b^\mu\tens \hat E_\kappa\tens \hat{\bar E}_{\dot\lambda}
		\]. 
		$\gamma_a$ can be constructed in an analogous fashion by using
		$(E_1,E_2,\bar E^1,\bar E^2)$ as basis of $\Delta_D$.
		Notice that in appendix \ref{sec:app:spinor-representation},
		we chose the Weyl representation as spinor representation of
		$\Clalg$. Though our theory does not depend on this choice
		(because all spinor representations of $\Clalg$ are
		equivalent), it has the benefit that the
		$\gamma$-tensor spinor (cf.\ appendix
		\ref{sec:app:gamma-tensor-spinor}) is given by 
		\[
		\hat\gamma_a = \left(
		\begin{array}{cc}
			0 & \hat\sigma_a^{\ A\dot X} \\
			\hat\sigma_{a\,\dot XA} & 0 
		\end{array}
		\right)
		\in \Gamma\left(T^*M \tens \GL\big( \mathscr{D}^{\halb,0} \oplus
		(\mathscr{D}^{0,\halb})^* \big) \right)
		\], 
		and thus enables easy transition between Dirac- and 2-spinor
		formalism. 

		Finally, it follows from the invariance of $\varepsilon$,
		$\sigma$ and $\gamma$ under simultaneous reference frame
		transformations (cf.\ appendix
		\ref{sec:app:reference-frame-transformations}) together with
		equation (O) above that these constructions do not depend on
		the choice of local section $F$. Thus, all these objects are
		well-defined \textit{globally.} Moreover we shall henceforth
		drop the hat on the bundle objects and just write
		$\varepsilon^{AB}$, $\sigma_a^{\ A\dot X}$ and $\gamma_a$. 

	\item It is well known that the spin structure $\mathscr{S}(M,g)$
		bears a connection naturally induced by the Levi-Civita
		connection on the connected frame bundle $\mathscr{F}_c(M,g)$.
		This connection induces covariant derivatives on all
		associated vector bundles (particularly, on all spinor
		bundles), which are again referred to as \textbf{Levi-Civita
		covariant derivatives,} and which are all \textbf{compatible}
		in the sense that they respect forming 
		tensor products, direct sums, dual bundles or complex
		conjugates of the bundles and on the level of representations.
		Therefore we may denote all Levi-Civita covariant derivatives
		on all the spinor bundles by $\nabla_a$ without ambiguity. 
		(In detail, this is elaborated in \onlinecite{Dipl}.)  

		Moreover, it can be shown that the fields $\sigma_a^{\ A\dot
		X}$, $\gamma_a$, $\varepsilon^{AB}$, $\varepsilon_{AB}$,
		$\varepsilon^{\dot X\dot Y}$ and $\varepsilon_{\dot X\dot Y}$
		on $M$ are all \textbf{parallel} with respect to the
		respective covariant derivatives, i.\,e.\ 
		\[
		\nabla_a\,\sigma_b^{\ A\dot X} \equiv 0
		\,,\qquad 
		\nabla_a\,\gamma_b \equiv 0
		\,,\qquad 
		\nabla_a\,\varepsilon^{AB} \equiv 0
		\]. 
		This is again a consequence of the invariance of $\sigma$,
		$\varepsilon$ and $\gamma$ under synchronous reference frame
		transformations, cf.\ appendix
		\ref{sec:app:reference-frame-transformations}. 
		Notice that this result admits \textbf{implicit index
		shifting} of spinor indices by means of $\varepsilon$, in the
		way described in appendix \ref{sec:app:index-notation}. 

	\item We remind that \textbf{index shifting} of spinor indices is done
		with respect to the anti-symmetric $\varepsilon$, as
		described in appendix \ref{sec:app:index-notation}. Moreover,
		for vector fields $X = X^a\in \Gamma(TM)$ and co-vector fields
		$\varphi_a\in\Gamma(T^*M)$, we use the notation 
		\[
		X^{A\dot X} := X^a\, \sigma_a^{\ A\dot X}
		\,,\qquad 
		\varphi_{A\dot X} := \alpha_a\, \sigma^{a}_{\ A\dot X}
		\]. 
		(Sometimes this is referred to as Dirac slash notation, though
		we will omit the slash.) In particular, for $\nabla_a$ we may
		write 
		\[
		\nabla_{A\dot X} = \sigma^{a}_{\ A\dot X} \nabla_a 
		\].
\end{enumerate}

\subsection{Differential operators and the Dirac equation on curved
spacetimes} 

\begin{enumerate}
	\item Let $\mathscr{E}$ be a general complex vector bundle on $(M,g)$
		with covariant derivative $\nabla^{\mathscr{E}}$. 
		For linear differential operators
		$\mathcal{P}\:\Gamma(\mathscr{E})\to\Gamma(\mathscr{E})$ on
		sections of $\mathscr{E}$ we generally adopt
		the terminology used in \cite{BaerGinPfae}. 
		Recall that for $\mathcal{P}$ of order $k$, the
		\textbf{principal symbol $s_\mathcal{P}$} is an object of the
		type 
		\[
		s_P \in \Gamma\left( (T^*M)^{\vee k}\tens \End(\mathcal{E})
		\right)
		\], 
		where $\vee k$ denotes the symmetrized $k$-fold tensor
		product. If $\mathcal{P}$ is of first order, the
		\textbf{principal part} (i.\,e.\ the highest
		order part) of $\mathcal{P}$ with respect to
		the covariant derivative $\nabla^{\mathscr{E}}$ may be written 
		as 
		\[
		\Phi\mapsto (s_\mathcal{P})^a\, \nabla^\mathcal{E}_a\, \Phi 
		\,,\qquad 
		\Phi\in\Gamma(\mathscr{E})
		\]. 
		(Notice, the principal symbol is independent of the covariant
		derivative $\nabla^\mathscr{E}$, while the principal part is
		not.) 

	\item A second order linear differential operator $\mathcal{L}\colon
		\Gamma(\mathscr{E}) \to \Gamma(\mathscr{E})$ is called
		\textbf{normally hyperbolic,} if its principal symbol
		$\sigma_L$ ``is given by the metric'': 
		\[
		\forall x\in M\,\forall \xi\in T_x^*M\colon \sigma_L(\xi,\xi)
		= g(\xi,\xi)\, \Id_{\mathscr{E}_x}
		\,,\]
		Equivalently this means that $L$ can locally be written as 
		\[
		L\Phi 
		= g^{\mu\nu}\frac{\partial^2}{\partial x^\mu\partial x^\nu}\Phi 
		+ \text{lower order derivatives of $\Phi$}
		\,,\quad \Phi\in\Gamma(\mathscr{E}) 
		\,.
		\]
		A \textit{first} order linear differential operator
		$\mathcal{P}\: \Gamma(\mathscr{E})\to \Gamma(\mathscr{E})$ is
		called \textbf{prenormally hyperbolic,} if there is another
		first order linear differential operator $\mathcal{Q}\:
		\Gamma(\mathscr{E})\to \Gamma(\mathscr{E})$, such
		that $\mathcal{PQ}$ is normally hyperbolic. 
		This concept was introduced in \cite{Mueh2011a} and is the
		central premise for the Cauchy problem of a first order
		operator $\mathcal{P}$ (on a globally hyperbolic manifold) to
		be solvable (i.\,e., well-posed). 

	\item The \textbf{Dirac operator} on the Dirac spinor
		bundle $\mathscr{D}^D$ is the first order linear differential
		operator $\mathcal{D}$ given by 
		\[
		\mathcal{D}\Psi = \gamma^a\,\nabla_a\Psi 
		\,,\qquad \Psi\in\Gamma(\mathscr{D}^D)
		\]. 
		The associated spin $\halb$ \textbf{Dirac field equation} for
		Dirac fields of mass $m\in\R$ reads: 
		\[
		0 = \mathcal{D}\Phi + \complexi\, m\,\Phi 
		\,,\qquad \Psi\in\Gamma(\mathscr{D}^D)
		\]. 
		Using the chiral decomposition $\mathscr{D}^D =
		\mathscr{D}^{\halb,0}\oplus (\mathscr{D}^{0,\halb})^*$, and
		writing $\Psi = \big( (\psi_1)^A, (\psi_2)_{\dot X} \big)^\tr$
		for
		$\psi_1\in\Gamma\big(\mathscr{D}^{\halb,0}\big)$ and $\psi_2\in
		\Gamma\big( (\mathscr{D}^{0,\halb})^* \big)$, we
		find 
		\[
		\begin{split}
		& 0 = \mathcal{D}\Psi + \complexi\,m\,\Psi 
		= \left(
		\begin{array}{cc}
			0 & \sigma_a^{\ A\dot X} \\
			\sigma_{a\,\dot XA} & 0 
		\end{array}
		\right) 
		\left(
		\begin{array}{c}
			\nabla_a\, (\psi_1)^A \\
			\nabla_a\, (\psi_2)_{\dot X}
		\end{array}
		\right)
		+ \complexi\, m\, 
		\left(
		\begin{array}{c}
			(\psi_1)^A \\
			(\psi_2)_{\dot X}
		\end{array}
		\right) \\
		\aequiv \quad & 
		\begin{cases} 
			0 = \nabla^{A\dot X}\, (\psi_2)_{\dot X} + \complexi\, m\,
			(\psi_1)^A\\
			0 = \nabla_{\dot XA}\, (\psi_1)^A + \complexi\, m\,
			(\psi_2)_{\dot X}
		\end{cases}
		\end{split}
		\]. 
		This shows how the Dirac equation decomposes into a system of
		the two coupled Weyl equations. 
\end{enumerate}

\newpage
\section{Fermionic quantum fields of higher spin}

\noindent 
It is our goal to investigate field equations for spinor fields of arbitrarily
high spin on our Lorentzian spacetime
manifold $(M,g)$. Loosely speaking, this could mean for fields of the general
form $\Psi^{A_1\ldots A_k\dot X_1\ldots\dot X_l}$, i.\,e.\ ``with a higher
number of dotted and undotted indices'', whence $\f{k+l}{2}$ will then be the
spin value. Such a field is called \textbf{Fermionic,} if the spin
$\f{k+l}{2}$ is half-integral. We shall specialize to Fermionic
fields later, but not from the beginning. 

Considering higher spin fields has a long history and there have been
various approaches, as pointed out in section \ref{sec:history}. In the works
of \nocite{Buchdahl1982a}\nocite{Buchdahl1982b}%
\textbf{Buchdahl} (1982a, 1982b) and %
\nocite{Wuensch1985}%
\textbf{Wünsch} (1985), fully symmetric generalized Dirac spinors 
\[
\Psi 
= \left( 
\begin{array}{l}
	(\psi_1)^{A A_1\ldots A_{k}}_{\qquad\quad \dot X_1\ldots \dot
	X_{l}} \\
	(\psi_2)^{\quad A_1\ldots A_{k}}_{\dot X\qquad\quad\dot X_1\ldots
	\dot X_{l}} 
\end{array}
\right) 
= \left( 
\begin{array}{l}
	(\psi_1)^{(A A_1\ldots A_{k})}_{\qquad\qquad (\dot X_1\ldots \dot
	X_{l})} \\
	(\psi_2)^{\quad (A_1\ldots A_{k})}_{(\dot X\quad\qquad\dot X_1\ldots
	\dot X_{l})} 
\end{array}
\right) 
\]
were considered, but in case of Wünsch, the principal symbol of his first
order operator is not invertible in normal direction (and thus the operator
fails to be prenormally hyperbolic), in case of Buchdahl the enforcement of
symmetry in all indices leads to additional and seemingly
unnatural constraint conditions (which is not a fatal problem
though). Moreover, in both cases, there doesn't seem to exist a canonical
Hermitian scalar product on the bundle with respect to which the differential
operators are not of positive definite type, and thus a canonical quantization
of the system using $\CAR$-algebras appears not to be feasible.  (Buchdahl's
and Wünsch's operators were studied in detail in \onlinecite{Dipl}). 

The \textbf{basic idea} of the present approach is to see in higher spin 
fields the structure of \textbf{twisted Dirac fields,} i.\,e.\ to consider
bundles of the type 
\[
\mathscr{M}^{\h{k},\h{l}} 
:= \mathscr{D}^D \tens \mathscr{\tilde D}^{\h{k},\h{l}}
= \left( \mathscr{D}^{\halb,0}\oplus (\mathscr{D}^{0,\halb})^* \right) \tens
\mathscr{\tilde D}^{\h{k},\h{l}} 
\,,\qquad k,l\in\set N
\]. 
Notice that the fibers of the bundle $\mathscr{M}^{\h{k},\h{l}}$ are
isomorphic to 
\[
\Delta_D\tens \tilde\Delta_{\h{k},\h{l}} 
= (\Delta_{\halb,0}\oplus\Delta_{0,\halb}^*)\tens \tilde\Delta_{\h{k},\h{l}}
\] 
and that the spin value of this representation is $\f{k+l+1}{2}$. 
Sections $\Psi\in\Gamma(\mathscr{M}^{\h{k},\h{l}})$ 
are of the form 
\[
\Psi = \left(
\begin{array}{l}
	(\psi_1)^{A A_1\ldots A_{k}}_{\qquad\quad\ \dot X_1\ldots \dot
	X_{l}} \\
	(\psi_2)^{\quad A_1\ldots A_{k}}_{\dot X\qquad\quad\dot X_1\ldots
	\dot X_{l}} 
\end{array}
\right) 
\].

\subsection{Twisted Prenormally Hyperbolic Operators} 

\noindent 
In this section we find as an auxiliary proposition that twists of prenormally
hyperbolic first order differential operators are again prenormally
hyperbolic. This general result will be applied to the higher spin operators 
introduced in the subsequent section. The concept of prenormal hyperbolicity
was introduced in \cite{Mueh2011a} as the central condition for the existence
of unique advanced and retarded Green's operators and of solutions to the
Cauchy problem for first order operators. 

Let $\mathscr{E}$ and $\mathscr{F}$ be complex vector bundles on our spacetime
manifold $(M,g)$, equipped with covariant derivatives $\nabla^\mathscr{E}$ and
$\nabla^\mathscr{F}$, respectively.  Let $\mathcal{P}\: \Gamma(\mathscr{E})
\to \Gamma(\mathscr{E})$ be a linear differential operator on $\mathscr{E}$.
Then the \textbf{twisted operator $\mathcal{P}^\mathscr{F}\:
\Gamma(\mathscr{E}\tens\mathscr{F})\to \Gamma(\mathscr{E}\tens\mathscr{F})$}
is the linear differential operator which is for simple tensors
$\psi\tens\alpha\in\Gamma(\mathscr{E}\tens\mathscr{F})$,
$\psi\in\Gamma(\mathscr{E})$, $\alpha\in\Gamma(\mathscr{F})$, given by 
\[
\mathcal{P}^\mathscr{F}(\psi\tens\alpha) 
:= \mathcal{P}\psi \tens \alpha+ (s_\mathcal{P})^a\psi\tens
\nabla^\mathscr{F}_a\alpha
\], 
where $s_\mathcal{P}\in\Gamma(T^*M\tens\End(\mathscr{E}))$ denotes the
principal symbol of $\mathcal{P}$. Notice that the principal part (i.\,e.\ the
highest order part) of $\mathcal{P}^\mathscr{F}$ is 
\[! \tag{$*$}
\psi\tens\alpha
\ \mapsto\ 
(s_\mathcal{P})^a \nabla^\mathscr{E}_a\psi\tens \alpha
+ (s_\mathcal{P})^a \psi\tens \nabla^\mathscr{F}_a\alpha
= \big( (s_\mathcal{P})^a \tens \Id_\mathscr{F} \big)\,
\nabla^{\mathscr{E}\tens\mathscr{F}}_a (\psi\tens \alpha)
\], 
where $\nabla^{\mathscr{E}\tens\mathscr{F}}$ denotes the induced tensor
product covariant derivative on $\mathscr{E}\tens\mathscr{F}$, given by 
\[
\nabla^{\mathscr{E}\tens\mathscr{F}}_a(\psi\tens\alpha) 
= \nabla^\mathscr{E}_a\psi\tens \alpha + \psi\tens \nabla^\mathscr{F}_a\alpha
\]. 
From $(*)$ we see that the principal symbol of the twisted operator
$\mathcal{P}^\mathscr{F}$ is given by 
\[
s_{\mathcal{P}^\mathscr{F}} = s_\mathcal{P}\tens\Id_\mathscr{F} 
\ \in\ 
\Gamma( T^*M\tens \End(\mathscr{E}\tens\mathscr{F}))
\]. 

\begin{proposition} \label{prop:twisted-prenorm-hyp}
	Let $\mathscr{E}$ and $\mathscr{F}$ be complex vector bundles on
	$(M,g)$.  Let $\mathcal{P}\:\Gamma(\mathscr{E})\to\Gamma(\mathscr{E})$
	be a prenormally hyperbolic differential operator of first order. Then
	the twisted operator $\mathcal{P}^\mathscr{F}$ on sections of
	$\mathscr{E}\tens\mathscr{F}$ is again prenormally hyperbolic.
\end{proposition}

\begin{proof}
	Let $\mathcal{Q}\:\Gamma(\mathscr{E})\to\Gamma(\mathscr{E})$ be a
	second first order operator such that $\mathcal{P},\mathcal{Q}$ form a
	complementary pair (terminology cf.\ \onlinecite[definition
	1]{Mueh2011a}). Then $\mathcal{P}\mathcal{Q}$ is of second order and
	normally hyperbolic, i.\,e.\ for all $\xi\in T^*M$, 
	\[
	s_{\mathcal{P}\mathcal{Q}}(\xi) 
	= s_\mathcal{P}(\xi)\,s_q(\xi) 
	= g(\xi,\xi)\,\Id_\mathscr{E}
	\]. 
	Notice that the twist $\mathcal{Q}^\mathscr{F}$ has principal symbol
	$s_{\mathcal{Q}^\mathscr{F}}(\xi) = s_\mathcal{Q}(\xi) \tens
	\Id_{\mathscr{F}}$ and thus we find 
	\[
	\begin{split} 
	s_{\mathcal{P}^\mathscr{F}\mathcal{Q}^\mathscr{F}}(\xi) 
	&= (s_\mathcal{P}(\xi) \tens \Id_{\mathscr{F}})\, 
	(s_\mathcal{Q}(\xi) \tens
	\Id_{\mathscr{F}}) \\
	&= (s_\mathcal{P}(\xi)\, s_\mathcal{Q}(\xi)) \tens
	\Id_{\mathscr{E}\tens\mathscr{F}} 
	= g(\xi,\xi) \, \Id_{\mathscr{E}\tens\mathscr{F}} \,.
	\end{split}
	\] 
	This means that $\mathcal{P}^\mathscr{F}\mathcal{Q}^\mathscr{F}$ is
	normally hyperbolic and thus, by definition of prenormal
	hyperbolicity, $\mathcal{P}^\mathscr{F}$ is prenormally hyperbolic. 
\end{proof}

\subsection{Higher spin quantum fields as twisted Dirac fields}

\noindent 
We now apply the general considerations above to the situation where
$\mathscr{E}$ is the Dirac spinor bundle, $\mathscr{E} = \mathscr{D}^D =
\mathscr{D}^{\halb,0}\oplus(\mathscr{D}^{0,\halb})^*$, equipped with the Dirac
operator 
\[
\mathcal{D}(\psi) = \gamma^a \nabla_a \psi
\,,\qquad 
\psi\in\Gamma(\mathscr{D}^D)
\]
(which is easily seen to be prenormally hyperbolic, cf.\
\onlinecite{Mueh2011a}). 
For $\mathscr{F}$ we may take any spinor bundle, i.\,e.\ arbitrary tensor
products and direct sums of $\mathscr{D}^{\halb,0}$ and
$(\mathscr{D}^{0,\halb})^*$. But as all of them can be broken down into direct
sums of symmetric spinors, we shall restrict our
attention here to $\mathscr{F} = \mathscr{\tilde D}^{\h{k},\h{l}}$, for yet
arbitrary $k,l\in\set N$. 

\begin{definition}[higher spin bundles and operators] \label{def:bundles}
	For $k,l\in\set N$, define the higher spinor bundles 
	\[
	\mathscr{M}^{k,l} 
	:= \mathscr{D}^D \tens \mathscr{\tilde D}^{\h{k},\h{l}}
	= \big(\mathscr{D}^{\halb,0} \oplus (\mathscr{D}^{0,\halb})^* \big)
	\tens \mathscr{\tilde D}^{\h{k},\h{l}}
	\] 
	and the higher spin first order differential operators 
	\[
	\mathcal{T}^{k,l} 
	:= \mathcal{D}^{\mathscr{\tilde D}^{\h{k},\h{l}}}
	\,,\qquad 
	\mathcal{T}^{k,l} \: \Gamma(\mathscr{M}^{k,l}) 
	\to \Gamma(\mathscr{M}^{k,l}) 
	\]. 
	(In words: $\mathcal{T}^{k,l}$ is the spin $\halb$ Dirac operator
	twisted by the spin $\f{k+l}{2}$ bundle $\mathscr{\tilde
	D}^{\h{k},\h{l}}$.) 
	Then for every $k,l\in\set N$ and for every physical mass $m\in\set R$
	we have the spin $s = \f{k+l+1}{2}$ field equation
	\[
	\mathcal{T}^{k,l}\Phi + \complexi\,m\,\Phi = 0 
	\,,\qquad \Phi\in\Gamma(\mathscr{M}^{k,l})
	\]. 
\end{definition}

\begin{remark}[principal symbol of $\mathcal{T}^{k,l}$]  
	The principal symbol of the Dirac operator $\mathcal{D}$ is
	$(s_\mathcal{D})^a = \gamma^a$. Thus, the principal symbol of
	$\mathcal{T}^{k,l}$ reads 
	\[
	(s_{\mathcal{T}^{k,l}})^a = \gamma^a \tens
	\Id_{\mathscr{\tilde D}^{\h{k},\h{l}}}
	\]
	and we can write 
	\[
	\mathcal{T}^{k,l}\Phi 
	= (\gamma^a\tens \Id_{\mathscr{\tilde D}^{\h{k},\h{l}}})\, \nabla_a
	\Phi 
	\overset{(2)}{=} \gamma^a\nabla_a\psi\tens\alpha + \gamma^a\psi \tens
	\nabla_a\alpha
	\], 
	while the second equality (2) holds only for simple tensors $\Phi =
	\psi\tens\alpha\in \Gamma(\mathscr{M}^{k,l})$, $\psi\in
	\Gamma(\mathscr{D}^D)$, $\alpha\in \Gamma(\mathscr{\tilde
	D}^{\h{k},\h{l}})$. 
	Here we denoted all
	covariant derivatives by $\nabla$, which is possible without ambiguity
	as the covariant derivatives on all 2-spinor bundles are compatible,
	cf.\ section \ref{sec:notation:spinors}.
\end{remark}

\begin{remark}[chiral decomposition of $\mathcal{T}^{k,l}$]
	The Dirac operator $\mathcal{D}$ has the chiral decomposition into
	positive and negative Weyl operators: 
	\[
	\mathcal{D} = \left(
	\begin{array}{cc}
		0 & \mathcal{D}_- \\
		\mathcal{D}_+ & 0 
	\end{array}
	\right)
	\,,\qquad 
	\begin{array}{l}
		\mathcal{D}_+\: \Gamma(\mathscr{D}^{\halb,0}) \to
		\Gamma\big( (\mathscr{D}^{0,\halb})^* \big) \,, \\
		\mathcal{D}_-\: \Gamma\big( (\mathscr{D}^{0,\halb})^* \big) 
		\to \Gamma(\mathscr{D}^{\halb,0}) \,.
	\end{array}
	\]
	This decomposition is inherited by the twisted Dirac operators
	$\mathcal{T}^{k,l}$: 
	\[
	\mathcal{T}^{k,l} = \left(
	\begin{array}{cc}
		0 & \mathcal{T}^{k,l}_- \\
		\mathcal{T}^{k,l}_+ & 0 
	\end{array}
	\right)
	\,,\qquad 
	\begin{array}{l}
		\mathcal{T}^{k,l}_+ =
		(\mathcal{D}_+)^{\mathscr{\tilde D}^{\h{k},\h{l}}}
		\,, \\
		\mathcal{T}^{k,l}_- =
		(\mathcal{D}_-)^{\mathscr{\tilde D}^{\h{k},\h{l}}}
		\,.
	\end{array}
	\]
\end{remark}

\noindent 
As in the case of the Dirac operator $\mathcal{D}$, 
the chiral decomposition of $\mathcal{T}^{k,l}$ is reflected in 2-spinor
notation of $\mathcal{T}^{k,l}$: 

\begin{remark}[higher spin fields in 2-spinor notation] 
	Notice that 
	\[
	\mathscr{M}^{k,l} 
	\isomorph \mathscr{D}^{\halb,0} \tens
	\mathscr{\tilde D}^{\h{k},\h{l}} \ 
	\oplus\  (\mathscr{D}^{0,\halb})^* \tens
	\mathscr{\tilde D}^{\h{k},\h{l}} 
	\]. 
	Thus, in 2-spinor notation, a section
	$\Phi\in\Gamma(\mathscr{M}^{\h{k},\h{l}})$ is written as 
	\[
	\Phi = \left(
	\begin{array}{l}
		(\varphi_1)^{AA_1\ldots A_k}_{\quad\qquad\dot X_1\ldots\dot
		X_l} \\
		(\varphi_2)^{\ \ A_1\ldots A_k}_{\dot X\ \qquad\dot
		X_1\ldots\dot X_l} 
	\end{array}
	\right)
	\,,\qquad
	\varphi_1\in\Gamma(\mathscr{D}^{\halb,0}\tens\mathscr{\tilde
	D}^{\h{k},\h{l}})
	,\ 
	\varphi_2\in\Gamma\big( (\mathscr{D}^{0,\halb})^* 
	\tens\mathscr{\tilde D}^{\h{k},\h{l}} \big)
	\], 
	and $\mathcal{T}^{k,l}\Phi$ is given by 
	\[
	\begin{split}
	\mathcal{T}^{k,l}\Phi 
	&= \underbrace{\left[ \left(
	\begin{array}{cc}
		0 & \sigma^{a\,A\dot X} \\
		\sigma^a_{\ \dot XA} & 0 
	\end{array}
	\right) \tens\Id_{\mathscr{\tilde D}^{\h{k},\h{l}}}
	\right]}_{\text{principal symbol $s_{\mathcal{T}^{k,l}}$}}
	\left(
	\begin{array}{l}
		\nabla_a (\varphi_1)^{AA_1\ldots A_k}_{\quad\qquad\dot
		X_1\ldots\dot X_l} \\
		\nabla_a (\varphi_2)^{\ \ A_1\ldots A_k}_{\dot X\ \qquad\dot
		X_1\ldots\dot X_l} 
	\end{array}
	\right) \\[3mm]
	&= \left(
	\begin{array}{l}
		\nabla^{A\dot X} (\varphi_2)^{\ \ A_1\ldots A_k}_{\dot
		X\ \qquad\dot X_1\ldots\dot X_l} \\
		\nabla_{\dot XA} (\varphi_1)^{AA_1\ldots A_k}_{\quad\qquad\dot
		X_1\ldots\dot X_l} 
	\end{array}
	\right) 
	= \left(
	\begin{array}{c}
		\mathcal{T}^{k,l}_- (\varphi_2) \\
		\mathcal{T}^{k,l}_+ (\varphi_1) \\
	\end{array}
	\right)
	\end{split}
	\] 
\end{remark}
	
\noindent 
Finally, as an immediate application of the general result proposition
\ref{prop:twisted-prenorm-hyp}, we may write down: 

\begin{theorem}[prenormal hyperbolicity, Green's functions and Cauchy
	problem] \label{thm:cauchy-problem}
	For every $k,l\in\set N$, we find for the higher spin first order
	differential operator $\mathcal{T}^{k,l}$ on sections
	of $\mathscr{M}^{k,l}$: 
	\begin{enumerate}[(a)]
		\item $\mathcal{T}^{k,l}$ is prenormally hyperbolic in the
			sense of \cite{Mueh2011a}. 

		\item There are unique advanced and retarded Green's operators
			(fundamental solutions) 
			\[
			G^{k,l}_\pm\: 
			\Gamma_0(\mathscr{M}^{k,l}) \to
			\Gamma(\mathscr{M}^{k,l})
			\] 
			for $\mathcal{T}^{k,l}$ (while $\Gamma_0$ denoted the
			space of compactly supported sections). 

		\item Let $\Sigma\subseteq M$ be a smooth spacelike Cauchy
			hypersurface and $m\in\R$ physical mass. 
			Then the Cauchy problem  
			\[
			(\mathcal{T}^{k,l})\quad 
			\begin{cases}
				\mathcal{T}^{k,l}\Phi + \complexi\,m\,\Phi = 0
				,\quad \Phi\in \Gamma(\mathscr{M}^{k,l}) \\
				\Phi|_\Sigma = \Phi_0
			\end{cases}
			\]
			has a unique solution for every initial datum
			$\Phi_0\in\Gamma_0(\mathscr{M}^{k,l}|_\Sigma)$.
			Moreover, this solution satisfies $\supp \Phi\subseteq
			J(\supp \Phi_0)$. 
	\end{enumerate}
\end{theorem}

\begin{proof}
	(a) is a corollary of proposition \ref{prop:twisted-prenorm-hyp}.
	After this is established, the general theory presented in
	\cite{Mueh2011a} can be applied (particularly, \onlinecite[theorems 1,
	2]{Mueh2011a}) to yield (b) and (c). 
\end{proof}

\subsection{Fermionic higher spin operators of indefinite type}
\label{sec:3c}

\noindent 
Let $(\mathscr{E}, \SProd*)$ be a Hermitian vector bundle on $(M,g)$, i.\,e.\
a complex vector bundle equipped with a Hermitian scalar product and
compatible covariant derivative $\nabla^\mathscr{E}$. A linear differential
operator $\mathcal{P}\: \Gamma(\mathscr{E})\to \Gamma(\mathscr{E})$ with
principal symbol $s_\mathcal{P}$ is called \textbf{of positive-definite
respectively of indefinite type with respect to $\SProd*$,}\footnote{%
	Cf.\ the terminology in the recent preprint \cite{BaerGin2011}.%
} if for every timelike, future directed co-vector
field $0\ne\xi\in \Gamma(T^*M)$, the sesqui-linear form\footnote{%
	Sesqui-linear means that the form is complex anti-linear in the first
	and linear in the second argument.}
\[
(\Psi,\Phi)_\xi := \SProd{\Psi}{\xi_a (s_\mathcal{P})^a \Phi} 
\,, \qquad \Psi,\Phi\in\Gamma(\mathscr{E})
\]
is (everywhere) positive-definite/indefinite, respectively. 

\begin{remark} \label{rem:selfadjoint}
	It is easy to see that $(\cdot,\cdot)_\xi$ is Hermitian if and only
	if $s_\mathcal{P}$ is self-adjoint with respect to $\SProd*$ (i.\,e.,
	$\All\xi\in\Gamma(T^*M) \all\Phi,\Psi\in\Gamma(\mathscr{E})\:
	\SProd{\Phi}{s_\mathcal{P}(\xi)\Psi} = \SProd{s_\mathcal{P}(\xi)
	\Phi}{\Psi}$). 
\end{remark}


\begin{example}[Dirac operator] \label{ex:dirac-positive-definite-type}
	It is well known that the spin $\halb$ Dirac operator $\mathcal{D}$ on
	$\mathscr{D}^D$ is of positive definite type with respect to the
	Hermitian scalar product on $\mathscr{D}^D$ given by pairing with the
	Dirac adjoint (cf.\ \onlinecite{Dimock1982}):

	The \textbf{Dirac adjoint} is the complex anti-linear mapping $\,^+\:
	\mathscr{D}^D\to(\mathscr{D}^D)^*$, given by 
	\[
	\Psi^+ := \twovec{\bar \varphi_{A}}{\bar\psi^{\dot X}} 
	\in \Gamma\left( (\mathscr{D}^D)^*\right)
	\quad\text{for}\quad
	\Psi = \twovec{\psi^A}{\varphi_{\dot X}}
	\in\Gamma(\mathscr{D}^D)
	\]. 
	By pairing spinors and co-spinors, this induces a
	Hermitian scalar product on $\mathscr{D}^D$: 
	\[
	\SProd{\Psi}{\Phi}_D 
	:= \Psi^+ ( \Phi) 
	\,,\quad \Psi,\Phi \in \Gamma(\mathscr{D}^D)
	\]. 
	With respect to this
	Hermitian scalar product, $\mathcal{D}$ is of positive-definite type.
	This means that for every nowhere vanishing, 
	timelike, future pointing co-vector field $\xi\in\Gamma(T^*M)$, 
	\[
	(\Psi,\Phi)_\xi 
	:= \SProd{\Psi}{\xi_a\gamma^a\Phi}_D
	= \Psi^+( \xi_a\gamma^a \Phi) 
	= (\psi_2)_A \xi^{A\dot X} (\varphi_2)_{\dot X} 
	+ (\psi_1)^{\dot X} \xi_{\dot XA} (\varphi_1)^A
	\]
	is positive definite in each fiber. (Here we used the notation $\Psi = 
	(\psi_1, \psi_2)^{tr}$, $\Phi = (\varphi_1, \varphi_2)^{tr} \in
	\Gamma(\mathscr{D}^D)$ and $\xi^{A\dot X} = \xi^a\sigma_a^{\ A\dot
	X}$.) 

	Moreover, it is easily seen that $\gamma^a$, the principal symbol of
	the Dirac operator, is self-adjoint with respect to $\SProd*_D$;
	hence, $(\cdot,\cdot)_\xi$ is also Hermitian. 
\end{example}

\noindent 
In order to generalize the Dirac spin $\halb$ case, we would 
like to have a Hermitian product $\SProd*_{\mathscr{M}^{k,l}}$ on our
higher spin bundles $\mathscr{M}^{k,l}$ with respect to which the operators
$\mathcal{T}^{k,l}$ are of positive-definite type. 
The differential operator being of positive-definite type with respect to the
Hermitian product on the bundle is the central prerequisite for our 
construction to qualify for \textbf{quantization using a $C^*$-algebra
representation of the canonical anti-commutation relations,} called
$\CAR$-algebra (see the well-known construction for the spin $\halb$ special
case by \onlinecite{Dimock1982} and the remarks on the general procedure in
\onlinecite[section V]{Mueh2011a}, as well as \onlinecite{Araki1970}). 

We shall now demonstrate that if $k=l$, there is a natural
Hermitian product on $\mathscr{M}^{k,l}$, but alas, the operators
$\mathcal{T}^{k,k}$ will be \textit{indefinite} with respect to it. Thus, we
will see that a quantization of the fields presented here is \textit{not}
possible in a straight forward fashion. Notice that setting $k=l$ is the point
where we restrict ourselves to \textbf{Fermionic fields,} as this 
implies that the spin $\f{k+l+1}{2}$ is half-integral. 

\medskip\noindent
The natural Hermitian product on the twisted bundle $\mathscr{M}^{k,k} =
\mathscr{D}^D \tens \mathscr{\tilde D}^{\h{k},\h{l}}$ will be the tensor
product of $\SProd*_D$ on $\mathscr{D}^D$ as declared in example
\ref{ex:dirac-positive-definite-type} and a Hermitian product $\SProd*_k$ on
$\mathscr{\tilde D}^{\h{k},\h{k}}$:

\begin{definition}[generalized Dirac adjoint and Hermitian product
	$\SProd*_{\mathscr{M}^{k,k}}$ on $\mathscr{M}^{k,k}$]%
	\label{def:hermitian-scalar-product}%
	Let two general spinors $\Phi,\Psi\in\Gamma(\mathscr{M}^{k,k})$ be
	written in 2-spinor notation as 
	\[
	\Phi 
	= \twovec{(\varphi_1)^{AA_1\ldots A_k}_{\quad\qquad\dot X_1\ldots\dot
	X_k}}%
	{(\varphi_2)^{\ \ A_1\ldots A_k}_{\dot X\ \qquad\dot X_1\ldots\dot X_k}}
	\,,\qquad 
	\Psi 
	= \twovec{(\psi_1)^{AA_1\ldots A_k}_{\quad\qquad\dot X_1\ldots\dot
	X_k}}%
	{(\psi_2)^{\ \ A_1\ldots A_k}_{\dot X\ \qquad\dot X_1\ldots\dot X_k}}
	\], 
	Then, on the Fermionic higher spinor bundle $\mathscr{M}^{k,k}$ we
	declare the Hermitian product 
	\[
	\SProd{\Phi}{\Psi}_{\mathscr{M}^{k,k}} := \Phi^+(\Psi) 
	\]. 
	Here, $\Phi^+$ is a \textbf{generalized concept of Dirac adjoint}
	co-spinor, defined as: 
	\[
	\Phi^+ := \left( 
	\begin{array}{l}
	(\bar\varphi_2)^{\ \ \dot X_1\ldots\dot X_k}_{A\ \qquad A_1\ldots A_k}
	\\
	(\bar\varphi_1)^{\dot X\dot X_1\ldots\dot X_k}_{\quad\qquad A_1\ldots
	A_k}
	\end{array}
	\right)
	\]. 
	Notice, this means that $\SProd*_{\mathscr{M}^{k,k}}$ is given by 
	\[
	\begin{split}
	\SProd{\Phi}{\Psi}_{\mathscr{M}^{k,k}} 
	&= (\bar\varphi_2)^{\ \ \dot X_1\ldots\dot X_k}_{A\ \qquad A_1\ldots
	A_k} (\psi_1)^{AA_1\ldots A_k}_{\quad\qquad\dot X_1\ldots\dot X_k} \\
	& \qquad\qquad 
	+ (\bar\varphi_1)^{\dot X\dot X_1\ldots\dot X_k}_{\quad\qquad
	A_1\ldots A_k}
	(\psi_2)^{\ \ A_1\ldots A_k}_{\dot X\ \qquad\dot X_1\ldots\dot X_k} 
	\,.
	\end{split}
	\]
\end{definition}

\noindent
Using the principal bundle $s_{\mathcal{T}^{k,k}}$ of $\mathcal{T}^{k,k}$, we
now construct the Hermitian product 
\[
(\Psi,\Phi)_\xi 
:= \SProd{\Psi}{s_{\mathcal{T}^{k,k}}(\xi)\, 
\Phi}_{\mathscr{M}^{k,k}}
\,, \qquad \Psi,\Phi\in\Gamma(\mathscr{M}^{k,k})
\] 
for every nowhere vanishing, future pointing timelike co-vector field
$\xi\in\Gamma(T^*M)$. In 2-spinor notation, this product reads: 
\[
\begin{split}
(\Phi,\Psi)_\xi
&= (\bar\varphi_2)^{\ \ \dot X_1\ldots\dot X_k}_{A\ \qquad
A_1\ldots A_k} \xi^{A\dot X}
(\psi_2)^{\ \ A_1\ldots A_k}_{\dot X\ \qquad\dot X_1\ldots\dot
X_k} \\
&\qquad\qquad 
+ (\bar\varphi_1)^{\dot X\dot X_1\ldots\dot X_k}_{\quad\qquad
A_1\ldots A_k} \xi_{\dot XA} 
(\psi_1)^{AA_1\ldots A_k}_{\quad\qquad\dot X_1\ldots\dot X_k}
\,.
\end{split}
\]

\begin{remark} \label{rem:hermitian}
	It is easily seen that the principal symbol $s_{\mathcal{T}^{k,k}} =
	\gamma\tens\Id_{\mathscr{\tilde D}^{\h{k},\h{k}}}$ of
	$\mathcal{T}^{k,k}$ is self-adjoint with respect to
	$\SProd*_{\mathscr{M}^{k,k}}$. Hence, $(\cdot,\cdot)_\xi$ is Hermitian
	(cf.\ remark \ref{rem:selfadjoint}).  
\end{remark}

\begin{remark}[indefiniteness of $(\cdot,\cdot)_\xi$] \label{rem:indefinite}
	For higher spin, i.\,e.\ for $k\ge 2$, the Hermitian product
	$(\cdot,\cdot)_\xi$ is indefinite and thus, $\mathcal{T}^{k,k}$ are of
	indefinite type with respect to $(\cdot,\cdot)_\xi$. 
\end{remark}

\begin{proof}
	It suffices to prove this in the fiber. 
	First, let $e_0,\ldots,e_3$ be
	the standard basis of $(\set R^4,\eta)$, let $E_1,E_2$ be the
	standard basis of $\Delta_{\halb,0}$, and set $\xi = e_0$. Then it is
	easily seen that for $\Phi = (\phi_1,\phi_2)^\tr$ with 
	\[
	\begin{split}
	&(\phi_1)^{AA_1\ldots A_k}_{\qquad\ \dot X_1\ldots\dot X_k}
	= (E_1 \tens\ldots\tens E_1)^{AA_1\ldots A_k} \,
	(\bar E_1^*\tens\ldots\tens\bar E_1^*)_{\dot X_1\ldots\dot X_k}
	\,, \\
	&(\phi_2)^{\ \ A_1\ldots A_k}_{\dot X\qquad\dot X_1\ldots\dot X_k}
	= (E_1 \tens\ldots\tens E_1)^{A_1\ldots A_k} \,
	(\bar E_1^*\tens\ldots\tens\bar E_1^*)_{\dot X\dot X_1\ldots\dot X_k} 	
	\,,
	\end{split}
	\]
	$(\Phi,\Phi)_\xi >0$. On the other hand, for $\Phi =
	(\phi_1,\phi_2)^\tr$ with 
	\[
	\begin{split}
	&(\psi_1)^{AA_1\ldots A_k}_{\qquad\ \dot X_1\ldots\dot X_k}
	= (E_1 \tens\ldots\tens E_1)^{AA_1\ldots A_k} \,
	(\bar E_2^*\tens\ldots\tens\bar E_2^*)_{\dot X_1\ldots\dot X_k} \\
	&\qquad\qquad\qquad \qquad\qquad
	- (E_1)^A (E_2 \tens\ldots\tens E_2)^{A_1\ldots A_k} \,
	(\bar E_1^*\tens\ldots\tens\bar E_1^*)_{\dot X_1\ldots\dot X_k}
	\,, \\
	&(\psi_2)^{\ \ A_1\ldots A_k}_{\dot X\qquad\dot X_1\ldots\dot X_k}
	= (E_1 \tens\ldots\tens E_1)^{A_1\ldots A_k} \,
	(\bar E_1^*)_{\dot X}\, (\bar E_2^*\tens\ldots\tens\bar E_2^*)_{\dot
	X\dot X_1\ldots\dot X_k}\\
	&\qquad\qquad\qquad \qquad\qquad
	- (E_2 \tens\ldots\tens E_2)^{A_1\ldots A_k} \,
	(\bar E_1^*\tens\ldots\tens\bar E_1^*)_{\dot X\dot X_1\ldots\dot X_k}
	\,,
	\end{split}
	\]
	we find $(\Psi,\Psi)_\xi < 0$. For $\xi$ different from $e_0$ it is
	not difficult to construct analogous examples: Just transform the
	basis $E_1,E_2$ by $S\in\SLtwo$ when transforming $\xi$ by
	$\Lambda(S)$. Then the same formulas as above will yield suitable
	examples. 
\end{proof}

\medskip\noindent 
This finding is of course fatal to the construction of a
$C^*$-algebra representation of the canonical anti-commutation relations
($\CAR$-algebra). However, compared to what we achieved in previous work (cf.\
\onlinecite{Dipl}), it is already a success that $(\cdot,\cdot)_\xi$ induces a
Hermitian (indefinite) product on the space of Solutions to the field
equation, $\mathcal{T}^{k,k} \Phi = 0$, which \textit{does not depend on
further choices.} This shall be demonstrated now: 

For the remainder of this section, fix an arbitrary $k\in\set N$. To simplify
notation, we set $\mathcal{T}:= \mathcal{T}^{k,k}$. 

\begin{enumerate}
	\item Let $\Sigma\subset M$ be a smooth spacelike Cauchy hypersurface.
		We define the \textbf{vector space of compactly supported
		Cauchy data} on $\Sigma$, 
		\[
		\mathscr{H}_\Sigma := \Gamma_0(\mathscr{M}^{k,k}|_\Sigma) 
		\], 
		on which we declare the product
		\[
		\SProd{\Phi_0}{\Psi_0}_\Sigma 
		:= \int\limits_\Sigma (\Phi_0, \Psi_0)_\mathfrak{n} 
		= \int\limits_\Sigma \mathfrak{n}^a
		\SProd{\Phi}{(s_\mathcal{T})_a \Psi}_{\mathscr{M}^{k,k}}
		\,,\qquad 
		\Phi_0,\Psi_0\in\mathscr{H}_\Sigma
		\], 
		where $\mathfrak{n}$ is the future pointing unit normal vector
		field along $\Sigma$. 
		$\SProd*_\Sigma$ is Hermitian by remark \ref{rem:hermitian}
		and for $k\ge2$ (i.\,e., for the higher spin case) indefinite
		by remark \ref{rem:indefinite}. 

	\item We define the \textbf{vector space of solutions with compactly
		supported Cauchy data:} 
		\[
		\mathscr{H} := \{ \Phi\in\Gamma(\mathscr{M}^{k,k}) \|
		\mathcal{T}^{k,k}\Phi = 0 \und
		\Phi|_\Sigma\in\mathscr{H}_\Sigma \}
		\]. 
		This is independent of the choice of $\Sigma$, which means
		that for a second smooth
		Cauchy hypersurface $\Sigma'\subset M$,
		$\supp(\Phi)\schnitt\Sigma'$ is again compact for
		$\Phi\in\mathscr{H}$. This is a consequence of global
		hyperbolicity of $(M,g)$, cf.\ \onlinecite[corollary
		1]{Mueh2011a}. 

		There is a canonical vector space isomorphism $\Xi_\Sigma\:
		\mathscr{H}_\Sigma \to \mathscr{H}$, given by assigning to
		$\Phi_0\in\mathscr{H}_\Sigma$ the unique solution to the
		Cauchy problem $(\mathcal{T}^{k,k})$ with initial datum
		$\Phi_0$. The inverse $\Xi_\Sigma^{-1}$ is given by
		the restriction map $\Phi\mapsto\Phi|_\Sigma$. 

		On $\mathscr{H}$ we obtain a Hermitian scalar
		product $\SProd*$ by pushing $\SProd*_\Sigma$ from
		$\mathscr{H}_\Sigma$ to $\mathscr{H}$ via $\Xi_\Sigma$: 
		\[
		\SProd{\Phi}{\Psi} 
		:= \SProd{\Phi|_\Sigma}{\Psi|_\Sigma}_\Sigma
		\,,\qquad \Phi,\Psi\in\mathscr{H}
		\]. 
		It is a crucial and non-trivial point that this construction
		is well-defined (i.\,e., independent of the choice of
		$\Sigma$). Thus, as central result of this section we shall
		prove: 
\end{enumerate}

\begin{theorem} \label{thm:4}
	The Hermitian product 
	$\SProd*$ on $\mathscr{H}$ is independent of the
	choice of smooth spacelike Cauchy hypersurface $\Sigma\subset M$. 
\end{theorem}

\noindent 
Notice, this result is well known for the spin $\halb$ special case of Dirac
spinor fields, cf.\ \cite{Dimock1982}, but so far it was not generalized to an
appropriate class of higher spin fields. 

\begin{proof} 
	Fix arbitrary $\Phi,\Psi\in\mathscr{H}$. We have to show 
	\[\tag{$*$}
	\int\limits_\Sigma \mathfrak{n}^a
	\SProd{\Phi}{(s_\mathcal{T})_a \Psi}_{\mathscr{M}^{k,k}} 
	= \int\limits_{\Sigma'} \mathfrak{n}'^a
	\SProd{\Phi}{(s_\mathcal{T})_a \Psi}_{\mathscr{M}^{k,k}}
	\] 
	for every second smooth Cauchy hypersurface $\Sigma'\subset M$ with
	future pointing unit normal vector field $\mathfrak{n}'$. 

	Declare the set $K := \supp(\Phi)\cup\supp(\Psi)$. It suffices to
	prove $(*)$ for the case that $K\cap\Sigma'$ lies completely in the
	future of $K\cap\Sigma$. Because if this is not the case,
	global hyperbolicity of $(M,g)$ always allows us to choose a third
	smooth spacelike Cauchy hypersurface $\Sigma''$ such that
	$K\cap\Sigma''$ lies in the future of the compact set
	$K\cap(\Sigma\cup\Sigma')$ (use a time-parameterized foliation of
	$(M,g)$ by Cauchy hypersurfaces, cf.\ \onlinecite{BerSan2005}).
	Transitivity would then yield equation $(*)$ for $\Sigma$ and
	$\Sigma'$. 

	Define the vector field $X^a\in\Gamma(TM)$, \[ X^a :=
	\SProd{\Phi}{(s_\mathcal{T})^a \Psi}_{\mathscr{M}^{k,k}} \], and
	notice that because $\mathcal{T}\Phi=0$ and $\mathcal{T}\Psi=0$, $X$
	is divergence free: 
	\[ 
	\begin{split} 
		\div X 
		= \nabla_a X^a 
		&=
		\SProd{\nabla_a\Phi}{(s_\mathcal{T})^a\Psi}_{\mathscr{M}^{k,
		k}} + \SProd{\Phi}{\nabla_a ( (s_\mathcal{T})^a \Psi)
		}_{\mathscr{M}^{k,k}} \\ 
		&=
		\SProd{(s_\mathcal{T})^a\nabla_a\Phi}{\Psi}_{\mathscr{M}^{k,k}}
		+
		\SProd{\Phi}{(s_\mathcal{T})^a\nabla_a\Psi}_{\mathscr{M}^{k,k}}
		\\ &= \SProd{\mathcal{T}\Phi}{\Psi}_{\mathscr{M}^{k,k}} +
		\SProd{\Phi}{\mathcal{T}\Psi}_{\mathscr{M}^{k,k}} \\ &= 0 
		\,.
	\end{split} 
	\] 
	Here, on the first summand we used self-adjointness of
	$s_\mathcal{T}$ (remark \ref{rem:selfadjoint}), for the second summand
	we used that $s_{\mathcal{T}} =
	\gamma\tens\Id_{\mathscr{\tilde D}^{\h{k},\h{k}}}$ is parallel (i.\,e.,
	$\nabla_a(\gamma\tens\Id_{\mathscr{\tilde D}^{\h{k},\h{k}}}) \equiv
	0$; for $\nabla_a\gamma\equiv 0$ cf.\ section
	\ref{sec:notation:spinors}).

	Choose $\Sigma'$ such that $K\cap\Sigma'$ lies completely in the
	future of $K\cap\Sigma$. Then due to global hyperbolicity, the set 
	$J_+( K\cap \Sigma ) \cap J_-( K\cap\Sigma')$ is compact (cf.\ e.\,g.\ 
	\onlinecite[lemma A.5.7]{BaerGinPfae}). 
	Choose $\Omega\subset M$ compact with piecewise smooth
	boundary $\pa\Omega$, such that the induced metric on the smooth part
	of $\pa\Omega$ is non-degenerate, and such that $J_+( K\cap \Sigma )
	\cap J_-( K\cap\Sigma') \subset \Omega$ and $K\cap\pa\Omega =
	K\cap(\Sigma\cup\Sigma')$. (In words, this means: $\Omega$ contains
	the compact set $J_+( K\cap \Sigma ) \cap J_-( K\cap\Sigma')$ and lies
	between $\Sigma$ and $\Sigma'$.)

	Let $\mathfrak{n}'$ be the future pointing unit normal vector field
	along $\Sigma'$ and let $\mathfrak{\tilde n}$ be the outward unit
	normal vector field along the smooth parts of $\pa\Omega$. 
	Notice that $\mathfrak{\tilde n}|_{\pa\Omega\cap K\cap\Sigma} =
	-\mathfrak{n}|_{\pa\Omega\cap\Sigma}$ and 
	$\mathfrak{\tilde n}|_{\pa\Omega\cap K\cap\Sigma'} =
	\mathfrak{n'}|_{\Omega\cap\Sigma'}$. 
	Then we finally obtain: 	
	\[
	\int\limits_{\Sigma'} \mathfrak{n}'^a X_a 
	- \int\limits_\Sigma \mathfrak{n}^aX_a 
	= \int\limits_{\pa\Omega} \mathfrak{\tilde n}^a\, X_a
	= \int\limits_\Omega \div X 
	= 0
	\], 
	where we used Gauss' divergence theorem for the second equality (in
	the version for semi-Riemannian manifolds, cf.\ e.\,g.\ 
	\onlinecite[theorem 1.3.16]{BaerGinPfae}). 
\end{proof}

\newpage
\subsection{General remarks on twisted prenormally hyperbolic operators} 

\noindent 
Notice that the proof of the previous theorem was not specific to our higher
spin construction but can literally be applied to the following
general setting: 

\begin{theorem}
	Let $(\mathscr{E}, \SProd*)$ be a Hermitian vector bundle with a
	compatible covariant derivative $\nabla$ and let
	$\mathcal{P}\: \Gamma(\mathscr{E})\to \Gamma(\mathscr{E})$ be a
	prenormally hyperbolic first oder differential operator whose
	principal symbol $s_\mathcal{P}$ is parallel with respect to the
	induced covariant 
	derivative on $T^*M\tens\End(\mathscr{E})$, and 
	self-adjoint with respect to $\SProd*$, i.\,e.\ 
	\[
	\All\xi\in\Gamma(T^*M) \all \Phi,\Psi\in\Gamma(\mathscr{E})\: 
	\SProd{\Phi}{s_\mathcal{P}(\xi) \Psi} 
	= \SProd{s_\mathcal{P}(\xi) \Phi}{\Psi} 
	\]. 
	Then for every nowhere vanishing, future pointing, timelike co-vector
	field $\xi\in\Gamma(T^*M)$, the sesqui-linear form 
	\[
	(\Phi,\Psi)_\xi := \SProd{\Phi}{s_\mathcal{P}(\xi)\Psi} 
	\,,\qquad \Phi,\Psi\in\Gamma(\mathscr{E})
	\]
	is Hermitian. For a smooth spacelike Cauchy hypersurface $\Sigma\subset
	M$ with future direct unit normal vector field $\mathfrak{n}$, the
	sesqui-linear form 
	\[
	\SProd{\Phi}{\Psi}
	:= \int\limits_\Sigma (\Phi|_\Sigma, \Psi|_\Sigma)_\mathfrak{n} 
	\] 
	on the space of
	solutions to $\mathcal{P}$ with compactly supported Cauchy data, 
	\[
	\mathscr{H} := \{ \Phi\in\Gamma(\mathscr{E}) \|
	\mathcal{P}\Phi = 0 \und
	\text{$\supp(\Phi|_\Sigma)$ is compact}\}
	\], 
	is well-defined (i.\,e.\ independent of the choice of Cauchy
	hypersurface $\Sigma$). 
	Moreover, $(\mathscr{H},\SProd*)$ forms a pre-Hilbert space (i.\,e.,
	$\SProd*$ is positive-definite), if $\mathcal{P}$ is of
	positive-definite with respect to $\SProd*$, and $\SProd*$ is
	indefinite if $\mathcal{P}$ is indefinite with respect to $\SProd*$. 
\end{theorem}

\begin{proof}
	For $(\cdot,\cdot)_\xi$ to be Hermitian, cf.\ remark
	\ref{rem:hermitian}. That the definition of $\mathscr{H}$ does not
	depend von $\Sigma$ is a consequence of \cite[corollary 1]{Mueh2011a}. 
	The proof that $\SProd*$ is well-defined is literally the proof of
	theorem \ref{thm:4} with $\mathcal{T}$ replaced by $\mathcal{P}$.
	Notice that compatibility of the covariant derivative on
	$\mathcal{E}$, parallelness and self-adjointness of $s_\mathcal{P}$
	are required there. 
\end{proof}

\noindent
In particular, this theorem applies to the following situation of a twisted 
pre-normally hyperbolic first order operator of positive-definite type: 

\begin{lemma} \label{lem:twist-of-positive-definite-type}
	Let $\mathcal{P}$ be a first order linear differential  operator on
	the Hermitian vector bundle $(\mathscr{E}, \SProd*_\mathscr{E})$ which
	is of positive definite type with respect to $\SProd*_\mathscr{E}$.
	Let $(\mathscr{F}, \SProd*_\mathscr{F})$ be a second Hermitian vector
	bundle with $\SProd*_\mathscr{F}$ positive definite. 
	Then the twisted operator $\mathcal{P}^\mathscr{F}$ is again of
	positive definite type with respect to the induced Hermitian scalar
	product $\SProd*_{\mathscr{E}\tens\mathscr{F}}$ on
	$\mathscr{E}\tens\mathscr{F}$.
\end{lemma}

\begin{proof}
	Recall that the induced product on $\mathscr{E}\tens\mathscr{F}$ 
	is on simple tensors $\psi\tens\alpha$ and $\varphi\tens\beta\in
	\Gamma(\mathscr{E}\tens\mathscr{F})$,
	$\psi,\varphi\in\Gamma(\mathscr{E})$,
	$\alpha,\beta\in\Gamma(\mathscr{F})$, given by 
	\[
	\SProd{\psi\tens\alpha}{\varphi\tens\beta}_{\mathscr{E}
	\tens\mathscr{F}} 
	= \SProd{\psi}{\varphi}_\mathscr{E} \SProd{\alpha}{\beta}_\mathscr{F}
	\]. 
	The twisted operator $\mathcal{P}^\mathscr{F}$ has principal symbol
	$(s_{\mathcal{P}^\mathscr{F}})^a = (s_\mathcal{P})^a\tens
	\Id_\mathscr{F}$ and thus we find for nowhere vanishing, timelike,
	future pointing $\xi\in \Gamma(T^*M)$: 
	\[
	\SProd{\psi\tens\alpha}{(\xi_a(s_\mathcal{P})^a \tens
	\Id_\mathscr{F})
	(\varphi\tens\beta)}_{\mathscr{E}\tens\mathscr{F}}
	= \SProd{\psi}{\xi_a(s_\mathcal{P})^a \varphi}_\mathscr{E}
	\SProd{\alpha}{\beta}_\mathscr{F} 
	\]. 
	This relation extends bilinearly to non-simple tensors and is
	positive definite as both factors are positive definite by assumption. 
\end{proof}
	
\noindent 
Twisting the Dirac operator by a spinor bundle belonging to an irreducible
spinor representation will not lead to a higher spin operator of
positive-definite type, as this would require a non-trivial irreducible
\textit{unitary} representation of $\SLtwo$, which does not exist. 
This is what we were taken in by in our construction above. However, this does
not already dismiss the program of searching for
a ``nice'' set of higher spin field equations, but rather calls for further
\textit{conceptual} clarification.

\newpage
\appendix

\section*{Appendix: Introduction to $\SLtwo$-spinor formalism}

\let\minsec=\subsection

\hiddensubsection{Clifford algebra of Minkowski vector space and its spinor
representation}
\label{sec:app:spinor-representation}

\noindent 
Let $(\set R^4,\eta)$ be standard \textbf{Minkowski vector space} with $\eta =
\mathrm{diag}(1, -1, -1, -1)$ and let $(\set R^4,\eta)_\C$ be its
complexification (i.\,e.\ the complex vector space $\set C^4$ equipped with
the Hermitian scalar product $\eta_\C = \eta\tens\C$ obtained by extending
$\eta$ complex anti-linearly in the first and linearly in the second argument). 
By $\big( (\R^4)^*, \eta_\C) \big)^*$ we denote the dual space equipped with
the induced inverse metric. 
Let $\Clalg := \Cl( (\R^4,\eta)_\C)$ be the complex \textbf{Clifford algebra}
of complexified Minkowski space (or, which is equivalent, $\Cl^c_{1,3} =
\Cl_{1,3}\tens\set C$).  It is well known that $\Clalg$ is isomorphic as $\set
C$-algebra to $\Mat_{4\times 4}(\C)$, the algebra of complex $4\times 4$
matrices. Such an isomorphism $\kappa\: \Clalg\to \Mat_{4\times 4}(\C)$
represents $\Clalg$ as matrices and is is called a
\textbf{spinor representation of $\Clalg$.} It is an important theorem that
all spinor representations of $\Clalg$ are equivalent. 

We call a collection of four complex $4\times 4$ matrices
$(\dirmat_0, \ldots, \dirmat_3)$ a \textbf{collection of Dirac matrices} with
respect to an ordered basis $(b_0, \ldots, b_3)$ of $\set R^4$, if 
\[! \label{dirac-matrix-relation}
\All\mu\nu\: \dirmat_\mu \dirmat_\nu + \dirmat_\nu\dirmat_\mu = 2\,\eta(b_\mu,
b_\nu) \cdot \set 1
\], 
where $\set 1$ denotes the $4\times 4$ identity matrix. 
Collections of Dirac matrices are intimately related to spinor representations
of $\Clalg$: From a spinor representation $\kappa\: \Clalg\to\Mat_{4\times
4}(\C)$ we obtain a collection of Dirac matrices with respect to a chosen
basis $(b_0,\ldots,b_3)$ by setting $\dirmat_\mu:= \kappa(b_\mu)$. 
Vice versa, if
a collection of Dirac matrices $(\dirmat_0,\ldots,\dirmat_3)$ with respect to a
basis $(b_0,\ldots,b_3)$ is given, we obtain a spinor representation $\kappa$
of $\Clalg$ by setting 
\[
\kappa(x) := x^\mu \dirmat_\mu
\,\quad\text{for vectors $x = x^\mu b_\mu\in\set R\subset \Clalg$}
\]. 
This is to be multiplicatively extended to all of $\Clalg$ and it follows
from the defining relations (\ref{dirac-matrix-relation}) that this yields a
well-defined representation of $\Clalg$. 

On the level of Dirac matrices, equivalence of all spinor representations of
$\Clalg$ is equivalent to the following two statements about \textbf{basis
transformations,} which are often useful: If $b_\mu\mapsto b'_\mu = A^\nu_{\
\mu} b_\nu$ for an invertible matrix $A^\nu_{\ \mu}\in\GL(4,\C)$ is a basis
transformation in $(\set R^4, \eta)$, then the matrices $\dirmat'_\mu :=
A^\nu_{\ \mu}\dirmat_\nu$ form a collection of Dirac matrices with respect to
the transformed basis $(b'_0, \ldots, b'_3)$, if $(\dirmat_0, \ldots,
\dirmat_3)$ is a collection of Dirac matrices with respect to $(b_0, \ldots,
b_3)$, as can be checked easily.  (I.\,e., the $\dirmat_\mu$ \textbf{transform
covariantly} under a basis transformation in $(\R^4,\eta)_\C$, and this is why
the index $\mu$ is written as subscript index.) Moreover, according to a
famous theorem by Pauli, if both $(\dirmat_0, \ldots, \dirmat_3)$ and
$(\dirmat'_0, \ldots, \dirmat'_3)$ are collections of Dirac matrices with
respect to a basis $(b_0, \ldots, b_3)$, there is an invertible $4\times 4$
matrix $S$, so that $\dirmat'_\mu = S\dirmat_\mu S^{-1}$ (basis transformation
in the representation space). This matrix $S$ is just the intertwining
isomorphism of the two spinor representations of $\Clalg$ induced by the two
different collections of Dirac matrices. 

This shows that \textbf{making a choice of Dirac matrices} in combination with
a choice of bases of $(\set R^4,\eta)$ comprises nothing more than the
fixation of one out of all the equivalent spinor representations of $\Clalg$.
Though all of our formulas will make use of an invariant notation, we will now
choose a certain collection of Dirac matrices and bases which is particularly
convenient for easily switching between Dirac 4-spinor notation and 2-spinor
notation. We are free to make this choice, our theory doesn't depend on
it, but we gain notational elegance. 

First, we declare the well-known $2\times2$ \textbf{Pauli spin matrices}
(extended by $\pauli_0$): 
\[
\pauli_0
= \left(
\begin{array}{rr}
	1 & 0 \\
	0 & 1
\end{array}
\right),\quad 
\pauli_1 
= \left( 
\begin{array}{rr}
0&1\\1&0
\end{array}
\right),\quad 
\pauli_2 = \left(\begin{array}{rr}
0&-i\\i&0
\end{array}
\right) , \quad
\pauli_3 = \left(
\begin{array}{rr}
1&0\\0&-1
\end{array}
\right)
\]. 
Recall that the following relations hold (as always, Latin indices $i,j,k$
take values $1,2,3$, Greek indices $\mu,\nu$ take values $0,\ldots,3$): 
\[
[\pauli_i,\pauli_j] = 2i\,\epsilon_{ijk}\,\pauli_k 
\,,\qquad 
\pauli_\mu^2 = \pauli_0
\,,\qquad 
\pauli_\mu^\dagger = \pauli_\mu^{-1} = \pauli_\mu
\]. 
Then it is easily checked that the following forms a collection of Dirac
matrices with respect to the \textbf{standard basis $(e_0,\ldots,e_3)$ of
$(\R^4,\eta)_\C$:}
\[
\dirmat_0 := \left(
\begin{array}{cc}
	0 & \pauli_0 \\
	\pauli_0 & 0
\end{array}
\right)
\,,\qquad 
\dirmat_i := \left(
\begin{array}{cc}
	0 & \pauli_i \\
	-\pauli_i & 0
\end{array}
\right)
\,,\qquad i = 1,2,3
\]. 
The spinor representation of $\Clalg$ induced by this collection of Dirac
matrices is commonly referred to as \textbf{Weyl representation.} From this
point on, we will always be using this representation, as performing concrete
calculations requires a concrete fixation of one out of all the equivalent
spinor representations of $\Clalg$. However, none of our results depends on
this choice, this choice just leads to notational benefits (which comes from
the fact that the Weyl representation already encodes a direct sum
decomposition of the Dirac spinor representation of $\Spin_0(1,3)$ into its
chiral parts, which enables easy switching between 4-spinor and 2-spinor
notation, see below).

\hiddensubsection{The $\Spin$-group of Minkowski vector space and $\SLtwo$}

\noindent 
Recall that every Clifford algebra $\Cl(V)$ for a real or complex vector space
$V$ carries a natural $\set Z_2$ grading, this is a direct sum decomposition
$\Cl(V) = \Cl^+(V) \oplus \Cl^-(V)$, where $\Cl^\pm(V)$ are the $\pm1$
eigenspaces of the grading automorphism $\alpha\: \Cl(V)\to\Cl(V)$, which is
given by multiplicative and additive extension of $\alpha(v) := -v$ for $v\in
V$. The $\Spin$ group of the Clifford algebra $\Cl^c_{1,3}$ is then defined as
$\Spin(1,3) := \Pin(1,3)\schnitt (\Cl_{1,3}^c)^+$, where $\Pin(1,3)\subset
\Cl^c_{1,3}$ is the multiplicative subgroup of $\Cl^c_{1,3}$ generated by the
subset $\{x\in \set C^4\| \eta_\C(x,x) = \pm 1\}$. It is well known that
$\Spin(1,3)$ is a \textit{real} 6-dimensional Lie group, and it is generated
by the bi-vectors 
\[
M_i := \epsilon_{ijk} \, e_je_k 
\,,\quad N_i := e_i e_0 
\,,\quad i,j,k = 1,2,3
\]. 
(This means that the $M_i$, $N_i$ form a basis of the real Lie-algebra
$\mathfrak{spin}(1,3)$ of $\Spin_0(1,3)$.) Here, $(e_0,\ldots,e_3)$ denotes
the standard basis of $(\R^4,\eta)_\C \subset \Clalg$. 
It is easily checked that amongst
these generators, the following commutator relations hold: 
\[! \tag{CR}
[M_i,M_j] = 2\, \epsilon_{ijk} \,M_k
\,,\quad 
[N_i,N_j] = -2\, \epsilon_{ijk}\, M_k 
\,,\quad 
[M_i, N_j] = 2\, \epsilon_{ijk}\, N_k
\]. 
The central role of $\Spin(1,3)$ for our considerations is due to the fact that
it is the natural universal covering group of the special orthogonal group
$\SO(1,3) = \mathcal{L}_+$, where $\mathcal{L}_+$ denotes the group of Lorentz
transformations with positive determinant. $\Spin(1,3)$ has two connected
components and we denote the connected component of unity by $\Spin_0(1,3)$.
It universally covers the proper orthochronous
Lorentz group $\mathcal{L}_+^\uparrow$ (Lorentz transformation without space
or time reversal). 

Finally, it is well known that $\Spin_0(1,3)$ is as real Lie
group isomorphic to $\SLtwo$, the group of unimodular complex $2\times 2$
matrices. As will become apparent below, spinor calculations are 
easier when dealing with elements of $\SLtwo$ (i.\,e., matrices) instead of
elements of $\Spin_0(1,3)$, and therefore we will now fix a concrete
isomorphism $\varphi\: \SLtwo\to\Spin_0(1,3)$. For $\SLtwo$ we have a set of
generators given by 
\[
m_i := -i\,\pauli_i
\,,\quad 
n_i := \pauli_i
\,,\quad 
i = 1,2,3
\]. 
It is easily checked that they fulfill the same commutator relations as (CR)
with capital letters replaced by lower case letters. Thus we obtain a Lie
algebra
isomorphism $d\varphi\: \mathfrak{spin}(1,3)\to\mathfrak{sl}(2,\C)$ (we denote
the Lie algebra of $\SLtwo$ by $\mathfrak{sl}(2,\C)$) by linear
extension of 
\[
\begin{split}
d\varphi\: m_i &\mapsto M_i  \,,\qquad i=1,2,3\\
n_i&\mapsto N_i \,.
\end{split}
\] 
As $\Spin_0(1,3)$ and $\SLtwo$ are simply connected, this integrates uniquely
to a Lie group isomorphism $\varphi\: \SLtwo\to\Spin_0(1,3)$.

\hiddensubsection{The Dirac spinor representation of $\SLtwo$ and its chiral
decomposition}

\noindent 
By $\Mat_{k\times k}(\C)$ we denote the algebra of complex $k\times k$
matrices. By $\GL(k,\C)$ we denote the subgroup of invertible $k\times k$
matrices. It is a Lie group with Lie algebra $\mathfrak{mat}_{4\times 4}(\C)$,
the space of complex $4\times 4$-matrices with matrix commutator. 

The restriction 
\[
D^D := \kappa|_{\Spin_0(1,3)}
\: \Spin_0(1,3) \to \GL(4,\C)
\]
of a spinor representation $\kappa\: \Cl^c_{1,3}\to \Mat_{4\times 4}(\C)$ to
the group $\Spin_0(1,3)\subset \Clalg$ is called \textbf{Dirac
spinor representation of $\Spin_0(1,3)$.} As all spinor
representations of $\Clalg$ are equivalent, a Dirac spinor representation of
$\Spin_0(1,3)$ is unique up to isomorphism. Let $dD^D\: \mathfrak{spin}(1,3)
\to \mathfrak{mat}_{4\times 4}(\C)$ denote the associated infinitesimal
representation of the Lie algebra $\mathfrak{spin}(1,3)$. Then with respect to
our concretely chosen spinor representation $\kappa$ (Weyl representation), we
find for the generators $M_i$ and $N_i$ of $\mathfrak{spin}(1,3)$: 
\[! \tag{GD}
dD^D(M_i) = \epsilon_{ijk}\, \dirmat_j\dirmat_k
= \left(
\begin{array}{cc}
	-i\,\pauli_i & 0 \\
	0 & -i\,\pauli_i 
\end{array}
\right) 
\,,\qquad 
dD^D(N_i) = \dirmat_i\dirmat_0
= \left(
\begin{array}{cc}
	\pauli_i & 0 \\
	0 & -\pauli_i 
\end{array}
\right) 
\]. 
All six generators of $D^D$ are of block diagonal form. Hence, the Dirac
spinor representation \textbf{splits into a direct sum} 
of two 2-dimensional \textbf{chiral subrepresentations.} 
For the first chiral subrepresentation, $D^+\: \Spin_0(1,3)\to \GL(2,\C)$, we
take the upper left block
of equations (GD), which means that, on the level of generators, $dD^+(M_i) =
-i\,\pauli_i$ and $dD^+(N_i) = \pauli_i$. Notice that by concatenation with
the isomorphism $\varphi\: \SLtwo\to\Spin_0(1,3)$ defined above, we obtain a
representation $D^+\circ\varphi$ of $\SLtwo$, infinitesimally given by 
\[
\begin{split}
	&d(D^+\circ\varphi)(m_i) 
	= (dD^+ \circ d\varphi)(m_i) 
	= -i\,\pauli_i 
	= m_i \,, \\
	&d(D^+\circ\varphi)(n_i) 
	= (dD^+ \circ d\varphi)(n_i) 
	= \pauli_i 
	= n_i \,.
\end{split}
\]
In words: The generators $m_i,n_i\in\mathfrak{sl}(2,\C)
\subset\mathfrak{mat}_{2\times 2}(\C)$ get represented by themselves. This
shows that \textbf{$D^+\circ\varphi$ is the defining representation of
$\SLtwo$,} i.\,e.\ $(D^+\circ\varphi)(S) = S$ for
$S\in\SLtwo\subset\GL(2,\C)$. This is the reason why it is much
more handy taking $\SLtwo$ instead of $\Spin_0(1,3)$ when doing calculations
with 2-spinors. 
We shall often drop the $\circ\varphi$ and
just write $D^+(S)$ and $D^D(S)$ for $S\in\SLtwo$. 

Recall that for a general representation of a Lie group $G$, $\rho\:
G\to\GL(k,\C)$, the \textbf{complex conjugate representation} $\bar\rho$ is
given by $\bar\rho(g) := \overline{\rho(g)}$, $g\in G$, and the
\textbf{dual representation} (contragredient representation), $\rho^*$, on the
dual space of the representation space, is given by $\rho^*(g) :=
(\rho(g^{-1}))^\tr$ (here, the superscript $\,^\tr$ means the transpose matrix).
On the Lie algebra level, this means for $\xi\in\mathfrak{g}$: $d(\rho^*)(\xi)
= - (d\rho(\xi))^\tr$. 
Now, combining both, we consider $\overline{D^+}^*$, and find for the
infinitesimal representation $d(\overline{D^+}^*)\:
\mathfrak{spin}(1,3)\to \mathfrak{mat}_{2\times 2}(\C)$: 
\[
\begin{split}
	&d(\overline{D^+}^*)(M_i) 
	= -\overline{d(D^+)(M_i)}^\tr 
	= -\overline{(-i\,\pauli_i)}^\tr
	= -i\, \pauli_i^\dagger 
	= -i\, \pauli_i \,, \\
	&d(\overline{D^+}^*)(N_i) 
	= -\overline{d(D^+)(N_i)}^\tr 
	= -\overline{\pauli_i}^\tr
	= - \pauli_i^\dagger 
	= - \pauli_i \,.
\end{split}
\]
Comparing this with the lower right blocks of equations (GD) and writing $D^-
:= \overline{D^+}$, this finally shows for the Dirac spinor representation
$D^D = \kappa|_{\Spin_0(1,3)}$: 
\[
D^D = D^+ \oplus (D^-)^* 
\]. 
This is the chiral decomposition of $D^D$, and the $D^\pm$ are called
\textbf{positive resp.\ negative Weyl spinor representation.} 
As shown above, as representations of $\SLtwo$, $D^+(S) = S$, for
$S\in\SLtwo$, is the defining
representation, $D^-(S) = \bar S$ is the complex conjugate of the defining
representation, and $(D^-)^*(S) = (\bar S^{-1})^\tr$. Considering the second
summand in the chiral decomposition of $D^D$ as dual representation makes it
possible to have $D^+$ and $D^-$ a complex conjugate pair of representations,
which will lead to an elegant notation of 2-spinors below (by the way, the
representations $D^-$ and $(D^-)^*$ are equivalent, as is easily seen using
the totally anti-symmetric $\epsilon$-spinor introduced below).

\hiddensubsection{Classification of finite dimensional $\SLtwo$ spinor representations} 
\label{sec:app:classification} 

\noindent 
If $D\: \SLtwo\to\GL(V)$ is a finite dimensional representation of $\SLtwo
\isomorph \Spin_0(1,3)$ on a complex vector space $V$, we call the elements of
$V$ \textbf{spinors of type $D$.} 
In particular, the spinors of type $D^D$ are called \textbf{Dirac spinors} or
\textbf{4-spinors} (as $D^D$ is 4-dimensional), and spinors of types $D_\pm$ are
called \textbf{positive/negative Weyl spinors.} Lets denote the representation
space of $D^D$ by $\Delta_D$ (of course, $\Delta_D = \set C^4$) and the
representation spaces of $D^\pm$ by $\Delta_\pm$ ($\Delta_\pm = \C^2$). 
As $D^D = D^+ \oplus (D^-)^*$, we have $\Delta_D = \Delta_+ \oplus \Delta_-^*$. 

It turns out that the two Weyl representation, $D^+$ and $D^-$, are not
equivalent and that they are (up to equivalence) the only irreducible
2-dimensional representations of $\SLtwo$. Moreover, they form a set of
\textbf{fundamental
representations of $\SLtwo$,} which means that every complex finite
dimensional \textbf{irreducible representation of $\SLtwo$ is isomorphic to a
symmetrized tensor product of $D^+$ and $D^-$,} i.\,e.\ to a representation of
the form 
\[
\repr{\h{k}}{\h{l}} := (D^+)^{\symtens k} \tens (D^-)^{\symtens l}
\quad\text{on}\quad 
\Delta_{\h{k},\h{l}} := (\Delta_+)^{\symtens k} \tens
(\Delta_-)^{\symtens l}
\,,\quad k,l\in\set N
\], 
were $X^{\symtens k}$ denotes the $k$-fold symmetrized tensor product 
$X^{\symtens k} := 
X\symtens \ldots \symtens X$ ($k$ times). The number $\f{k}{2}+\f{l}{2}$ will
be interpreted as the physical particle spin. 
For the dimensions of these representations it is easily checked that 
\[
\dim \repr{\h{k}}{\h{l}} = (2k+1)(2l+1)
\]. 

Notice that in this systematic notation, $D^+ = \repr{\halb}{0}$ and $D^- =
\repr{0}{\halb}$. Moreover, as $D^- = \bar D^+$, we find that complex
conjugation just switches $k$ and $l$: 
\[
\overline{\repr{\h{k}}{\h{l}}} \isomorph \repr{\h{l}}{\h{k}}
\quad\text{on}\quad 
\overline{\Delta_{\h{k},\h{l}}} \isomorph \Delta_{\h{l},\h{k}}
\]. 
It is important to bear in mind that only fully symmetrized tensor products of
irreducible spinor representations are again irreducible. This can also be
seen from the \textbf{Clebsch-Gordon formula,} which reads in our setting: 
\[
\repr{\h{k}}{\h{l}} \tens \repr{\h{k'}}{\h{l'}} 
\isomorph \bigoplus_{i = |k-k'|}^{k+k'}\ \bigoplus_{j = |l-l'|}^{l+l'}
\repr{\h{i}}{\h{j}}
\]. 
As an important special case, we have: 
\[
\repr{\halb}{0} \tens \repr{\h{k}}{\h{l}} 
\isomorph \repr{\h{k+1}}{\h{l}} \oplus \repr{\h{k-1}}{\h{l}}
\].

\hiddensubsection{Index notation for 2-spinors} 
\label{sec:app:index-notation} 

\noindent 
Though they are both 2-dimensional complex vector spaces, it is important to
formally distinguish the representation spaces $\Delta_{\halb,0}$ and
$\Delta_{0\halb}$. This is why in index notation we use \textbf{undotted capital
indices} $A$, $B$, \ldots, $X$, $Y$, \ldots, for positive, and \textbf{dotted
capital indices} $\dot A$, $\dot B$, \ldots, $\dot X$, $\dot Y$, \ldots\ for
negative Weyl spinors: 
\[
\psi = \psi^A\in\Delta_{\halb,0} 
\quad\text{and}\quad
\varphi = \varphi^{\dot X}\in\Delta_{0,\halb}
\]. 
As usual, co-spinor indices are subscripted: 
\[
\psi = \psi_A\in\Delta_{\halb,0}^*
\quad\text{and}\quad
\varphi = \varphi_{\dot X}\in\Delta_{0,\halb}^*
\]. 
A mixed (higher) spinor
$\Psi\in\Delta_{\h{k},\h{l}}$ will then be denoted 
\[
\Psi = \Psi^{A_1\ldots A_k\dot X_1\ldots\dot X_l}\in \Delta_{\h{k},\h{l}}
\]. 
All these indices are considered \textbf{abstract indices,} 
i.\,e.\ they are \textit{not} numerical variables taking
values (for such we would use Greek letters $\mu$, $\nu$, \ldots), rather,
they are abstract labels. The main benefit from abstract index notation is that
it maintains the formal aspects of the Einstein summation convention and of
index shifting (isomorphic mapping between vector space and its dual), 
but still \textbf{all expressions are invariant,} they do not assume a choice of
reference basis.\footnote{%
	Notice: An expression holds in
	abstract index notation if and only if \textit{for every choice of
	reference basis,} it holds in component index notation. 
	Thus, also the reader not familiar with abstract index notation can
	follow this document if he just assumes his favorite choice of basis
	(or, later, local reference frame) chosen.}

The transformation of spinors under an $S\in\SLtwo$ reads in this
notation: 
\[
\begin{split}
&\repr{\halb}{0}(S)\psi = S^{A}_{\ B}\, \psi^B 
\,,\quad 
\repr{0}{\halb}(S)\varphi = \bar S^{\dot X}_{\ \dot Y}\, \varphi^{\dot Y}
\,,\quad \\
&\repr{\h{k}}{\h{l}}(S)\Psi 
= S^{A_1}_{\ \ B_1}\cdots S^{A_k}_{\ \ B_k} \,
\bar S^{\dot X_1}_{\ \ \dot Y_1}\cdots \bar S^{\dot X_l}_{\ \ \dot Y_l} \,
\Psi^{B_1\ldots B_k\dot Y_1\ldots\dot Y_l}
\,.
\end{split}
\]
We are using \textbf{implicit contraction} (analog to implicit summation in
classical component index notation), but this may only be taken
across indices which are either both dotted or both undotted, and only
diagonally (one index upper, the other lower, i.\,e.\ one spinor, the other
co-spinor). 

Once again, notice that spinors in the irreducible representation
$\Delta_{\h{k},\h{l}}$ are fully symmetric. This means that a $\Psi\in
\Delta_{\halb,0}^{\tens k}\tens\Delta_{0,\halb}^{\tens l}$ is an element of
the \textit{smaller} space $\Delta_{\h{k},\h{l}}$, if and only if 
\[
\Psi^{A_1\ldots A_k\dot X_1\ldots\dot X_l}
= \Psi^{(A_1\ldots A_k)(X_1\ldots\dot X_l)}
\]. 
Here, the brackets are the \textbf{symmetrization operator:} 
\[
\Phi^{(A_1\ldots A_k)} := \f{1}{k!} \sum_{\pi\in S_k} \Phi^{A_{\pi(1)}\ldots
A_{\pi(k)}}
\], 
where $S_k$ is the group of permutations of numbers $1,\ldots,k$. This
definition holds analogously for symmetrization on dotted indices.

As it is a 2-dimensional space, $\Delta_{\halb,0} = \Delta_+ = \C^2$ carries a
unique (up to a scalar factor) symplectic structure $\varepsilon_{AB}$, i.\,e.\
an anti-symmetric bilinear form. We fix this form by declaring that with
respect to the standard basis of $\C^2$ it has the component representation 
\[
\varepsilon_{\mu\nu} = \left(
\begin{array}{cc}
	0 & 1 \\
	-1 & 0
\end{array}
\right)
\]. 
We call this \textbf{the $\varepsilon$-spinor.} On the co-spinor space
$\Delta_+^*$, we declare $\varepsilon^{AB}$ such that
$\varepsilon^{AB}\varepsilon_{CB} = \Id^{A}_{\ C}$, where $\Id$ denotes the
identity matrix. By complex conjugation we obtain $\varepsilon_{\dot X\dot Y}$
on $\Delta_{0,\halb} = \Delta_-$ and $\varepsilon^{\dot X\dot Y}$. 
It is easily checked that $\varepsilon_{AB}$ is invariant under
$D^{\halb,0}\tens D^{\halb,0}$ (analogously for $\varepsilon^{AB}$,
$\varepsilon_{\dot X\dot Y}$ and $\varepsilon^{\dot X\dot Y}$. 

Notice, we didn't designate a scalar product on $\Delta_+$ and $\Delta_-$
with respect to which we might shift indices (i.\,e.\ use the induced
isomorphism between space and dual space). \textbf{Index shifting} in 2-spinor
formalism though will be done with respect to the anti-symmetric
$\varepsilon$-spinors: \textbf{Lowering with the first, raising with the second
index of $\varepsilon$.} We have: 
\[
\begin{split}
&\psi_B = \varepsilon_{AB}\,\psi^A 
\,,\qquad 
&&\varphi_{\dot Y} = \varepsilon_{\dot X\dot Y}\,\varphi^{\dot X}  \\
&\psi^A = \varepsilon^{AB}\, \psi_B
\,,\qquad 
&&\varphi^{\dot X} = \varepsilon_{\dot X\dot Y}\,\varphi_{\dot Y}
\end{split}
\]

\hiddensubsection{$\sigma$-tensor spinor and vector representation}
\label{sec:app:sigma-tensor-spinor}

\noindent 
We declare \textbf{the $\sigma$-tensor spinor,} which is a mixed object
$\sigma_a^{\ A\dot X} \in \set (\R_\C^4)^*\tens \Delta_{\halb,\halb}$, where
$\R_\C^4 = \C^4$ denotes complexified Minkowski vector space. With respect to
the standard bases in $(\set R^4, \eta)_\C$ and
$\Delta_\pm$, it is given by the Pauli matrices: 
\[
[\sigma_\mu^{\ \kappa\dot\lambda}] := \f{1}{\sqrt{2}}\, \pauli_\mu
\,,\qquad \mu=0,\ldots,3
\]. 
($\kappa=$ row index, $\dot\lambda=$ column index). 
$\sigma_a^{\ A\dot X}$ can be considered as a map 
$\sigma\: (\R^4,\eta)_\C \to \Delta_{\halb,\halb}$, given for $x =
x^a\in\R_\C^4$ by 
\[
\sigma(x) 
= x^a \sigma_a^{\ A\dot X} 
=: x^{A\dot X}
\]. 
It is easy to see that $\sigma$ is an isomorphism and that the inverse 
$\sigma^{-1}\: \Delta_{\halb,\halb}\to(\R^4,\eta)_\C$ is given by $\sigma^a_{\
A\dot X}$, where all indices are shifted according to the rules (i.\,e., using 
$\varepsilon_{AB}$, $\varepsilon_{\dot X\dot Y}$ and $\eta^{ab}$). 

The $\sigma$-spinor plays a central role in the
formalism because it can be shown that for
every $S\in\SLtwo$, the mapping $\lambda(S)\: (\R^4,\eta)_{\C}\to
(\R^4,\eta)_{\C}$, defined by the diagram 
\[
\begindc{\commdiag}[3]
	\obj(0,0)[a]{$\qquad\qquad$}
	\obj(-5,0)[a']{$\Delta_{\halb,\halb}\ni x^{A\dot X}$}
	\obj(60,0)[b]{$\qquad\qquad\qquad$}
	\obj(68,0)[b']{$S^A_{\ B}\,\bar S^{\dot X}_{\ \dot Y}\,
	x^{B\dot Y} \in \Delta_{\halb,\halb}$}
	\obj(0,20)[c]{$\quad$}
	\obj(-8,20)[c']{$(\set R^4,\eta)_\C \ni x^a$}
	\obj(60,20)[d]{$\qquad\qquad$}
	\obj(67,20)[d']{$\lambda(S)x\in (\set R^4,\eta)_\C$}
	\mor{a}{b}{$D^{(\halb,\halb)}(S)$}[1,6]
	\mor{c}{a}{$\sigma$}[1,6]
	\mor{d}{b}{$\sigma$}[1,6]
	\mor{c}{d}{$\lambda(S)$}[1,6]
\enddc
\]
is a restricted, orthochronous Lorentz transformation. $\lambda$ as defined by
commutativity of this diagram is the \textbf{universal covering map $\lambda\:
\SLtwo\to\restlor$} and, as the commutative diagram tells us, the
representation of $\SLtwo$ through $\lambda$ on $(\R^4,\eta)_\C$ is
\textit{equivalent}
to $D^{(\halb,\halb)}$ on $\Delta_{\halb,\halb}$, with $\sigma$ as
intertwiner! 

This is the reason why the $\SLtwo$-representation $D^{(\halb,\halb)}$ is also
referred to as \textbf{vector representation} (notice, it has spin 1). 
Every vector $x=x^a\in(\R^4,\eta)_\C$ naturally corresponds to a 
$D^{(\halb,\halb)}$-spinor 
\[
x^{A\dot X}  := x^a \, \sigma_a^{\ A\dot X}
\]. 
The internal consistency of this formalism gets apparent one more time when we
consider the push-forward of $\eta_{ab}$ through $\sigma$: 
\[
\eta_{A\dot XB\dot Y} = \eta_{ab}\, \sigma^a_{\ A\dot X}\,\sigma^b_{\ B\dot Y}
\]. 
An easy calculation reveals: 
\[
\eta_{A\dot XB\dot Y} = \varepsilon_{AB}\varepsilon_{\dot X\dot Y}
\]. 
This means: The canonical \textit{anti-symmetric} structures $\varepsilon$ on
$\Delta_{\halb,0}$ and $\Delta_{0,\halb}$ multiply together to form the
\textit{symmetric} Lorentzian scalar product $\eta_\C$ on $(\R^4,\eta)_\C$.

\hiddensubsection{The $\gamma$-tensor spinor}
\label{sec:app:gamma-tensor-spinor}

\noindent 
Recall that the spinor representation $\kappa$ represents $\Clalg$ as matrices
on $\C^4$. This representation was fixed by our choice of Dirac matrices
$\dirmat_\mu$. Restricting $\kappa$ to $\Spin_0(1,3)$, we called the
representation space $\C^4 =: \Delta_D$ and introduced the chiral decomposition
$\Delta_D = \Delta_{\halb,0}\oplus \Delta_{0,\halb}^*$. Let a Dirac
spinor $\Psi\in\Delta_D$ be given with respect to this chiral decomposition as 
\[
\Psi = \left(
\begin{array}{c}
	(\psi_1)^A \\
	(\psi_2)_{\dot X}
\end{array}
\right)
\,,\qquad 
\psi_1 = (\psi_1)^A \in \Delta_{\halb,0}
\,,\quad 
\psi_2 = (\psi_2)_{\dot X} \in \Delta_{0,\halb}^*
\]. 
Let a vector $x = x^a\in(\R^4,\eta)_\C\subset \Clalg$ be given in components
$x^\mu$ with respect to the standard basis $(e_0,\ldots,e_3)$. Then we find by
application of all the definitions we made: 
\[
\kappa(x)\Psi 
= x^\mu\,\dirmat_\mu \Psi 
= \left(
\begin{array}{cc}
	0 & x^\mu\,\pauli_\mu \\
	\eta_{\mu\mu}\, x^\mu\, \pauli_\mu 
\end{array}
\right) \left(
\begin{array}{c}
	\psi_1 \\
	\psi_2
\end{array}
\right)
= \sqrt{2}\, \left(
\begin{array}{c}
	x^a\, \sigma_a^{\ A\dot X}\, (\psi_2)_{\dot X} \\
	x^a\, \sigma_{a\,\dot X A}\, (\psi_1)^A
\end{array}
\right)
\]. 
Thus, the Dirac matrices $\dirmat_\mu$ with respect to the standard
basis $(e_0,\ldots,e_3)$ give rise to the so called \textbf{$\gamma$-tensor
spinor} $\gamma_a\in (\R^4,\eta)_\C^* \tens \GL( \Delta_D)$, which might as
well be considered a map $\gamma\: (\R^4,\eta)_\C\to \GL( \Delta_D )$. 
Accounting for the chiral decomposition $\Delta_D = \Delta_{\halb,0}\oplus
\Delta_{0,\halb}^*$, our choice of Dirac matrices (the Weyl representation,
cf.\ section \ref{sec:app:spinor-representation}) 
induces $\gamma_a$ to be given by: 
\[
\gamma_a = \left(
\begin{array}{cc}
	0 & \sigma_a^{\ A\dot X} \\
	\sigma_{a\, \dot XA} & 0
\end{array}
\right)
\]. 
(Notice, in this notation, the sign change of $\eta_{\mu\mu}\,\pauli_\mu$
for $\mu=1,2,3$ in the lower left corner is implemented implicitly when 
the indices $A\dot X$ are shifted down
by means of $\varepsilon_{AB}\varepsilon_{\dot X\dot Y}$.) This is the reason
we prefer the Weyl representation. It allows easy transition between Dirac
spinor notation and 2-spinor notation.

\hiddensubsection{Simultaneous reference frame transformation and invariance of
$\varepsilon$, $\sigma$ and $\gamma$}
\label{sec:app:reference-frame-transformations}

\noindent
``2-spinors are objects that change sign under a rotation by $2\pi$''---this
is what is often written in text books. It just says that in general, a
Lorentz transformation does not contain all the information about a
\textbf{reference frame transformation} of a physical system which is
described by elements of the Minkowski vector space $(\R^4,\eta)$ and spinors
of type $D$, where $D$ is a general finite-dimensional representation of
$\SLtwo$ on a complex vector space $\Delta$. Rather, such a transformation is
fully qualified only by giving an element of $\SLtwo$: 

Let a reference frame of our system by given by bases $(b_0,\ldots,b_3)$ of
$(\set R^4,\eta)$ and $(B_1,\ldots,B_k)$ of $\Delta$ ($k := \dim\Delta$). 
Moreover, let $\lambda\:\SLtwo\to\restlor$ denote the universal covering map
as given above. 	
Then an element $S\in\SLtwo$ induces a reference frame transformation
\[
b_\mu \mapsto b'_\mu := \lambda(S)(b_\mu)  
\quad  \text{and} \quad 
B_i \mapsto B'_i := D(S)(B_i) \,,\quad i = 1,2
\]. 
Now, as a very useful property of the formalism it can be shown that the
objects $\varepsilon^{AB}$, $\sigma_a^{\ A\dot X}$ and $\gamma_a$ are
\textbf{invariant} under such simultaneous reference frame transformations!
For example in the case of $\sigma_a^{\ A\dot X}$, this means that the
component representation of $\sigma_a^{\ A\dot X}$ with respect to bases
$(b^0,\ldots,b^3)$ of $(\R^4,\eta)_\C^*$, $(B_1,B_2)$ of $\Delta_{\halb,0}$
and $(\bar C_1,\bar C_2)$ of $\Delta_{0,\halb}$ remains untouched when these
bases are simultaneously transformed under $\lambda(S)^*$, $D^{(\halb,0)}(S)$
and $D^{(0,\halb)}(S)$ for any $S\in\SLtwo$. Analog statements hold for
$\varepsilon^{AB}$, $\varepsilon_{AB}$, $\varepsilon^{\dot X\dot Y}$,
$\varepsilon_{\dot X\dot Y}$ and $\gamma_a$.

\nocite{Verch2001}
\nocite{BrunettiFredenhagenVerch2003}

\bibliography{cauchy-problem.bib}

\end{document}